Valentina BOEVA

U900 Institut Curie/INSERM/Mines ParisTech
"Bioinformatics and Computational Systems Biology of Cancer"
Institut Curie, rue d'Ulm 26, 75248, Paris

# DECIPHERING REGULATION IN EUKARYOTIC CELL: FROM SEQUENCE TO FUNCTION

Contents

# ABSTRACT

For the last 8 years, a transversal topic of my research has been the development and application of computational methods for DNA sequence analysis. The methods I have been developing aim at improving our understanding of the regulation processes happening in normal and cancer cells. This topic connects together the projects presented in this thesis.

Two chapters of the thesis represent major areas of my research interests: (1) methods for deciphering transcriptional regulation and their application to answer specific biological questions, and (2) methods to study the genome structure and their application in cancer studies.

The first chapter predominantly focuses on transcriptional regulation. Here I describe my contribution to the development of methodology for the discovery of transcription factor binding sites and the positioning of histone proteins. I also explain how sequence analysis, in combination with gene expression data, can allow the identification of direct target genes of a transcription factor under study, as well as the physical mechanisms of its action. As two examples, I provide the results of my study of transcriptional regulation by (i) oncogenic protein EWS-FLI1 in Ewing sarcoma and (ii) oncogenic transcription factor Spi-1/PU.1 in erythroleukemia.

In the second chapter, I describe the sequence analysis methods aimed at the identification of the genomic rearrangements in species with existing reference genome. I explain how the developed methodology can be applied to detect the structure of cancer genomes. I provide an example of how such an analysis of tumor genomes can result in a discovery of a new phenomenon: chromothripsis, when hundreds of chromosomal rearrangements occur in a single cellular catastrophe.

The thesis is concluded by listing the major challenges in high-throughput sequencing analysis. I also discuss the current top questions demanding the integration of sequencing data.

# RESUMÉ [FRENCH]

*La régulation dans la cellule eucaryote à déchiffrer : de séquence à fonction.*

Lors des 8 dernières années, un thème transversal de ma recherche a été le développement et l'application de méthodes computationnelles à l'analyse des séquences d'ADN. Les méthodes que je développe visent à améliorer notre compréhension des processus de régulation qui se déroulent dans les cellules normales et cancéreuses. Les projets présentés dans ce mémoire s'inscrivent dans ce thème.

Les deux chapitres du mémoire constituent les principaux domaines de mes intérêts de recherche: (1) les méthodes pour déchiffrer la régulation transcriptionnelle et leur application pour répondre à des questions biologiques spécifiques, et (2) des méthodes pour étudier la structure du génome et leurs applications dans les études sur le cancer. Le premier chapitre parle principalement de la régulation transcriptionnelle. Je décris ici ma contribution à l'élaboration d'une méthodologie pour la découverte de sites de liaison de facteurs de transcription et le positionnement des histones. J'explique également comment une analyse de séquences, en combinaison avec des données d'expression des gènes, peut permettre l'identification de gènes cibles directs d'un facteur de transcription étudié, ainsi que des mécanismes physiques de cette action. En tant qu'exemples, je fournis les résultats de mon étude sur la régulation transcriptionnelle par (i) la protéine oncogénique EWS-FLI1 dans le sarcome d'Ewing et (ii) le facteur de transcription oncogénique Spi-1/PU.1 dans les érythroleucémies.

Dans le deuxième chapitre, je décris les méthodes d'analyse de séquences pour l'identification des réarrangements génomiques chez des espèces chez lesquelles il existe un génome de référence. J'explique comment la méthodologie développée peut être appliquée à la détection de la structure du génome tumoral. J'explique comment une telle analyse des génomes tumoraux peut entraîner une découverte d'un phénomène nouveau: chromothripsis, où des centaines de réarrangements chromosomiques se produisent dans une seule et unique catastrophe cellulaire.

Dans la conclusion du mémoire je parle des principaux défis dans l'analyse de séquençage à haut débit. Je discute aussi les questions actuelles exigeant l'intégration des données de séquençage.

# PREFACE

*To my beloved husband Valentine, my family and friends*

From the beginning of my carrier, the main subject of my research has been analysis of sequence data with the goal to reveal the link between DNA sequence and its function.

This interest started with the analysis of motif patterns in DNA sequences. I studied divergent tandem repeats during my PhD and transcription factor binding motifs during my first postdoctoral fellowship. Later on, I started working on assembly, clustering and statistical analysis of small genomic sequences – reads – coming from high-throughput DNA sequencing. Mappings of reads to the genome of reference can provide different pieces of information about the molecular events in the cell. For instance, the reads can be source of information about the exact location of DNA binding proteins (ChIP sequencing), provide information about gene expression (RNA sequencing) and mutations (genome resequencing). I have worked with these three applications of the sequencing technology. I have developed analysis methods that are nowadays used all over the world, for instance, by such institutions as Inserm, BC Cancer Agency, Broad Institut and NIH.

It is difficult to separate the methodological work I lead from the biological questions I answer. Most of the time, a biological phenomenon I studied was bringing me the development of novel data analysis methods. I was often among the first researchers in the world to work on a given type of data (ChIP-seq, cancer genome resequencing) while methods necessary for the appropriate data analysis had not been developed yet. In the two chapters of this thesis, I first speak about the methods and then provide the results of the research projects that used the analysis of the corresponding data type. To illustrate the application of ChIP-seq analysis methods, I describe the projects aimed at the discovery of mechanisms of transcriptional regulation by oncogenic transcription factors PU.1/Spi-1 in erythroleukemia and EWS-FLI1 in Ewing sarcoma. To provide an example of analysis of genome resequencing data, I present the results of the study of the rearrangements in the neuroblastoma genome and demonstrate the discovery of chromothripsis in this type of cancer.

# 1 INTRODUCTION

Almost each cell of our organism carries a recording media containing the information about when and how to assemble the "worker" molecules of our body – proteins. This recording media is called genome. Not every part of genome has equal importance. The consequences of DNA damage of some parts can never show up, while the damage in other parts can lead, under favorable environmental circumstances, to the changes in cell phenotype, cell death or cancer. The latter regions are called *functionally* important. Sequence analysis plays a pivotal role in the annotation of functionally important regions of the genome.

There is common metaphor to compare the human genome with a 6 billion letter book, were words and sentences (although not separated by spacers and commas) are different functional elements of the genome: coding regions, distal regulatory elements (enhancers), protein binding sites, repeat sequences, etc. Volumes of this book are chromosomes, and chapters are gene loci, i.e. instructions on how to make proteins. There are about 20-25 thousand of such chapters in the human genome. Unlike books written by humans, the genome book contains some chapters to be read from left to right and others be read from right to left. The genomic regions called promoters contain guidelines about the orientation of the reading (transcription).

In 2001, the International Human Genome Sequencing Consortium reported the first draft sequence of the human genome (Lander et al. 2001). Since then, a set of amendments have been done to improve the human genome reference sequence (fill gaps between contigs and resolve the structure of repetitive sequences). However, as if having a text in Ancient Greek does not allows a English speaking person to understand what it is about (even if the latter is a mathematician and knows the Greek alphabet), having an assembled genome does not imply one can understand the meaning of each particular genomic sentence. Therefore, it is too early to say that we have "decoded" the human genome. What we have done: we assembled the book, but we still don't know all elements of the language it was written in.

Currently the great community effort is being made to understand the logic of the genomic language. As in linguistics, we are interested to study *syntax* (how the language is formed, its structure), *semantics* (how meaning is inferred from the words) and *pragmatics* (how meaning is inferred from the context).

> *Considering the genome as a book written in a foreign language and making parallels with linguistics, we can study the meaning of the genomic text at two distinctive levels: semantic level (how meaning is inferred from the words) and pragmatic level (how meaning is inferred from the context).*
>
> *The study of genomic semantics involves functional annotation of genomic subsequences: coding regions, non-coding RNAs and regulatory regions (promoters, TFBS and binding sites of regulatory RNAs).*
>
> *The study of genomic pragmatics concerns the analysis of chromatin structure (i.e. genomic "context"). Chromatin properties such as presence of repressive or active chromatin, and chromatin 3D structure have direct influence on the accessibility of TFBSs and gene promoters to the transcriptional machinery of the cell.*

To describe the *syntax* of the genome, we can say that the genome of each human individual is written in 4-letter alphabet and it includes short and long repetitive sequences. In the genome, some *k*-mer words occur much more frequently than others; some *k*-mers are depleted, e.g. CG-pairs are depleted compared to other dinucleotides. The genomes of two human individuals are more than 99% similar. The difference is mostly due to the highly variable positions (single nucleotide polymorphisms, SNPs) or copy number polymorphisms (T. 1000 G. P. Consortium 2010; Conrad et al. 2010). Given that each genome is slightly different from another, the question that we now start to raise is how to represent the unified reference genome while taking into account the common genomic variants.

The understanding of genomic *semantics* includes the detection of the functional units in the genomic text and comprehension of how their combination is translated into the meaning, i.e. which genomic parts may be included into the RNA transcript and what may be the corresponding transcription rate. Speaking about functional units, we can list coding regions (exons) interspersed with long nonconding regions (introns), 5' and 3' transcribed but untranslated regions (UTRs), promoters and enhancers. The latter are most frequently located upstream promoters and inside introns (Andersson et al. 2014). In addition, there are non-coding RNA loci. For instance, microRNA loci often form clusters preceded by one common promoter region. Long non-coding RNA loci sometimes overlap coding regions, thereby giving an example of how the same text can be read in different directions providing different meanings.

We annotated more or less all exons of the human genome, and using the CAGE technique (Shiraki et al. 2003) we detected more or less all transcription start sites (TSSs, FANTOM projects). To map the enhancers of the human genome, several research groups took the coordinated initiative to locate TFBSs in dozens of cell types (the ENCODE project and FANTOM consortium)

(T. E. P. Consortium 2012; Andersson et al. 2014). Enhancers are often tissue specific, implying they are functional in one cell context and not in the other. They are composed of transcription factor binding sites (TFBSs), which are, all over again, functional in the cell-type specific manner. The sequence analysis can be applied to annotate TFBSs and understand how their combination (including the distance between them, their order and orientation) leads to the modulation of gene expression. The similar type of analysis is done to map miRNA binding in 3' UTRs. Here again, the combination of miRNA binding sites defines, in a cell-type dependent manner, the translation rate of the corresponding mRNA transcript.

Even more challenging than to uncover the genome semantics is to comprehend the genome *pragmatics*, i.e. to understand the ways in which the context contributes to the meaning. The context in the genome is defined in large part by the chromatin landscape. Chromatin consists of DNA and DNA associated proteins. DNA is tightly bound around histone octamers (nucleosomes), which form the skeleton of chromatin. Histones carry dozens of posttranslational modifications that can be recognized by specific proteins (Berger 2007). In addition, specific histone variants can be employed in regions bearing particular function such as active and poised promoters: (H2A.Z) (Ku et al. 2012) or centromere borders (H3.3 phosphorylated on serine 31) (Hake et al. 2005). Chromatin also interacts with structural proteins, e.g. proteins of nuclear lamina. It was shown that genes located in lamina associated domains (LADs) of chromatin tend to be silent, or expressed at undetectable or very low levels (Kind et al. 2013).

One can think about chromatin landscape as of the second layer of information over the DNA code. If we think about the genome as of a book, then chromatin represents marginal notes across the pages. For example, these notes can tell the reader, i.e. the cell transcriptional machinery, to skip a chapter (repressive chromatin) or to prepare a chapter to be read (poised transcription defined by chromatin bivalent domains) (Ku et al. 2008). We can define at least five principal chromatin types using information about chromatin proteins binding DNA and key histone modifications (Filion et al. 2010). It was demonstrated that the same promoter coupled to a reporter gene can show 1000 fold different transcriptional activity when inserted in the "active" or "inactive" chromatin environment (Akhtar et al. 2013). This observation indicates that the pages of the genomic book are sometimes "glued" together that makes the whole their content inaccessible to the transcriptional machinery. The histone profiles and the binding site maps for a variety of DNA-associated proteins for a large number of cell types are in the process of being created by the research community (ENCODE project). Studying these maps will allow us to get closer to the understanding of the transcription semantics.

In the last decade, novel methods such as Chromosome Confirmation Capture (3C), its extensions 4C, 5C and hi-C, and Chromatin Interaction Analysis by Paired-End Tag Sequencing (ChIA-PET), have been applied to map interacting DNA regions and get insights into the 3D structure of the chromatin (Ernst 2012). Using these techniques, the smallest units of the spatial genomic compartmentalization have been discovered. These units make 200-kilobase to 1 Mb in size and are often called topologically associating domains (TADs) (Nora et al. 2012). TADs are characterized by numerous long-range interactions of loci in the same domain and much more sporadic interactions of loci in adjacent domains. TADs are usually delimited by binding sites of transcriptional repressor CTCF, which is known to work as chromatin barrier, blocking the interaction between enhancers and promoters. The further study of the chromatin interaction maps will reveal the role of chromatin in genome regulation in normal cells and disease.

In cancer cells, genomic mutations and rearrangements introduce changes in the functional meaning of the genomic text both on the level of semantics (DNA) and pragmatics (chromatin).

Point mutations and indels can modify the protein coding regions; such changes are likely to be reflected in the structure of the corresponding proteins. Also, small size mutations in promoters and enhancers can result in the destruction of TFBSs and thus can bring alterations in the program of gene expression. Large deletions and amplifications can remove or duplicate one or more genes. This may lead to significant changes in the gene expression rates. "Chimeras" may be formed at the places of fusion of two genes. Such chimeric genes encode proteins that couple two existing protein functions or, in rare cases, possess a novel function in the cell. Also, translocations and other large structural variants can disrupt the structure of LADs and TADs, and thus result in aberrant gene expression. Using recent genome sequencing techniques, we can now characterize these events in cancer (Medvedev, Stanciu, and Brudno 2009; Medvedev et al. 2010; Boeva, Popova, et al. 2012; Koboldt et al. 2012).

Changes of the chromatin context, also called epigenetic changes, have been shown to play a role in cancer development and progression (Esteller 2007). One example is recently discovered phenomenon of long range epigenetic silencing (LRES). In LRES, long genomic regions including dozens of genes are simultaneously silenced by chromatin modifications (placement of repressive chromatin marks such as H3K27me and DNA methylation). These regions often include several tumor suppressor genes whose expression becomes turned off in a concerted manner in cancer. LRES has been shown to affect gene expression in many cancer types including bladder cancer (Stransky et al. 2006), colorectal cancer (Dallosso et al. 2012; Frigola et al. 2006), breast cancer (Novak et al. 2008) and prostate cancer (Coolen et al. 2010). By using the analysis of high-throughput sequencing data (ChIP sequencing), we can now identity epigenetic changes in cancer (Ashoor et al. 2013).

# 2     PATTERNS IN THE GENOMIC SEQUENCE AND THEIR ROLE IN REGULATION OF GENE EXPRESSION

The human genome contains a variety of motif hits or patterns (Figure 1). A far from exhaustive list of genomic patterns include (i) tandem repeats and transposable elements, (ii) stretches of GC- or AT- rich sequences (e.g. CpG islands), (iii) binding sites of DNA associated proteins, (iv) splice sites, and (v) DNA and RNA binding sites of non-coding RNA molecules. Different pattern often overlap each other or share function. Since we perform functional annotation of RNA binding sites using the genomic sequence as the system of reference, I deliberately put together binding sites located on DNA and RNA.

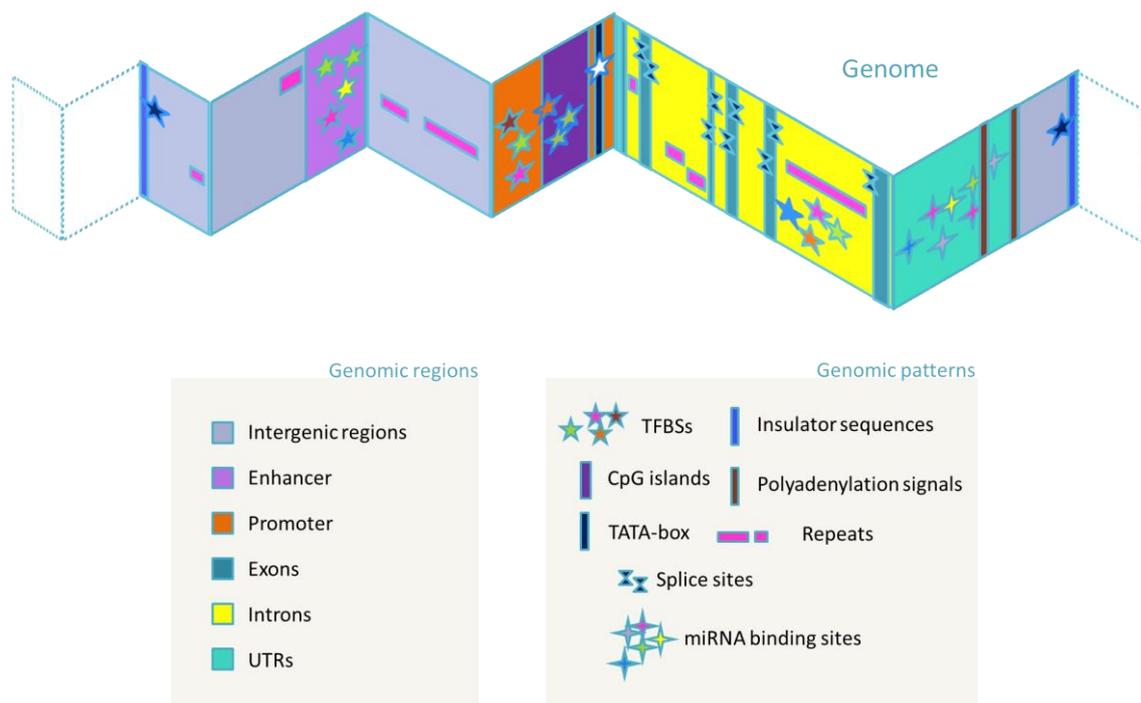

**Figure 1**. Illustration of genomic patterns (out of scale). Different functional groups of genomic elements (exons, regulatory modules, etc.) are shown by vertical stripes. Different genomic patterns (repeats including tandem repeats and transposable genetic elements, binding sites of transcription factors, splice sites, miRNA binding sites) are shown by colored shapes.

**Repeats**   More than 45% of the human genome corresponds to repetitive sequences (Derrien et al. 2012). Among them, one distinguishes tandem repeats (DNA is repeated in head to tail fashion: microsatellites, minisatellites and satellite sequences) and interspersed repeats (similar sequences are located throughout the genome). The latter correspond to such as transposable elements as SINEs and LINEs, accounting for 12.5% and 20% of the human genome, respectively.

The category of tandem repeats accounts for 10 to 15% of the human genome. While short tandem repeats can serve as binding sites for specific transcription factors (Shi et al. 2000; Guillon et al. 2009), long satellite repeats can play a role in the 3D structure shaping of the genome. For instance, the α-satellite family of repeats (~ 171 bp tandem repeats) are bound by the fundamental component of the centromere CENP-C and are essential for centromere function by ensuring proper chromosome segregation in mitosis and meiosis (Politi et al. 2002).

**AT- or GC- rich sequences**     AT- or GC- rich sequences are often located in the gene promoters playing a role in transcription initiation. Approximately 24% of human genes contain an AT-rich sequence within the core promoter, with 10% containing a canonical TATA-box motif (TATAWAWR, W=A/T, R=A/G) (Yang et al. 2007). The TATA-box recruits the TATA binding protein (TBP), which unwinds the DNA; also, due to weaker base-stacking interactions among A and T (rather than G and C), the AT-rich sequence facilitates the unwinding. The remaining 76% of the human promoters are GC-rich and contain multiple binding sites of the transcriptional activator SP1 (Yang et al. 2007).. As much as 56% of the human genes, including most of the housekeeping genes, possess CpG islands, i.e. 300-3000bp GC-rich sequences around gene TSS with the high density of CpG dinucleotides. The high methylation level of CpG sites in CpG islands is associated with transcriptional repression. Polycomb group (PcG) repressor proteins recognize CpG islands that are unmethylated and unprotected by transcription factors (Klose et al. 2013). PcG proteins associate with DNA methyltransferases responsible for methylation of CpG islands (Viré et al. 2006). Also, some components of the PcG proteins have histone methyltransferase activity and trimethylate histone H3 on lysine 27, which is a mark of transcriptionally silent chromatin.

**Transcription factor binding sites**     Transcription factors are proteins with DNA binding activity involved into the regulation of transcription. They are able to modulate gene expression by binding to the promoter or enhancer sequences. Although these proteins have DNA binding domains, in some cases they may bind the DNA indirectly, by interacting with another transcription factor. For instance, PU.1 and GATA-1 (transcription factors playing a critical role in the differentiation of hematopoietic lineages) interact through the ETS domain of PU.1 and the C-terminal finger region of transcription factor GATA-1; as a result, PU.1 can bind to DNA both directly and indirectly, through the assistance of GATA-1 (Figure 2) (Burda, Laslo, and Stopka 2010). Transcription factors have preferences for a set of DNA sequences called "binding motifs". Sequences forming the motif have different binding affinity. There are several ways to define binding motifs. One of the most common ways is the positional weight matrix (PWM) containing the log odds weights for computing the binding affinity score. This topic is discussed in details in the next section. As it will be shown later, in some cases the same transcription factor is able to bind quite dissimilar motifs.

It is interesting that in such situation the choice of motif bound may define the action of this transcription factor on gene expression (Guillon et al. 2009).

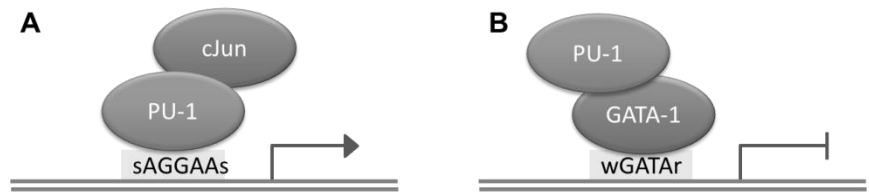

**Figure 2**. Direct and indirect binding of transcription factor PU.1 to DNA. **A**. Direct binding of transcription factor PU-1 to DNA to the consensus motif sAGGAAs, which may lead to transcriptional activation. **B**. Indirect binding of PU.1 to DNA, which may lead to transcriptional repression.

Transcription factors often interact with each other or compete for DNA binding. Then, their binding sites co-localize or overlap (Jie Wang et al. 2012). Co-localization of TFBSs can be also due to the communal action of a set of transcription factors: First, transcription factors capable of binding inactive chromatin bind to DNA and create the open chromatin environment through the recruitment of histone acetyltransferases. Then, other transcription factors (lacking the above capability) become able to bind DNA and activate gene transcription by interacting with the RNA polymerase machinery (Farnham 2009). Analysis of the distance and orientation preferences between the sites of co-binding transcription factors helps to predict possible protein-protein interactions and enables discovery of the mechanisms of transcriptional regulation by transcription factors.

**Splice sites** Splicing is the event when introns are removed from the pre-messenger RNA transcript. The remaining exons are joined together to form later on mature messenger RNA. Generally in eukaryotes, the process of splicing is catalyzed by spliceosomes. These proteins recognize a donor site (almost invariantly GU at the 5' end of the intron), a branch site (an adenine nucleotide followed by pyrimidine-rich tract near the 3' end of the intron) and an acceptor site (almost invariantly AG at the 3' end of the intron) on RNA transcript. DNA mutation of a splice site may have a wide range of functional consequences, among them exclusion of an exon from the mature mRNA, inclusion of an intron or a part of an intron. The latter often results in a disruption of the reading frame or a premature stop codon, and thus gives rise to a defective or truncated protein.

**miRNA binding sites** While the binding of regulatory proteins to promoter and enhancer DNA regions regulates the expression of the targeted protein at the level of transcription, in animals the binding of micro RNA molecules (miRNAs) to the 3'UTR region of mRNA transcript regulates the amount of the protein at the post-transcriptional level. The interaction of a miRNA as part of an active RNA-induced silencing complex (RISC) with a 3'UTR of the targeted mRNA transcript results in either inhibition of translation or increased degradation of this transcript. miRNA complex recognizes the 6-8 nucleotides at the mRNA 3'UTR which is complementary to the miRNA "seed" region (Bartel 2009). In the human genome there are more than 2000 unique miRNAs. One miRNA can target several genes, and the same 3'UTR can be targeted by multiple miRNAs. Sequence analysis of gene's 3'UTR coupled with the analysis of evolutionary conservation of the gene 3'UTR allows predicting miRNA-target pairs (Yue, Liu, and Huang 2009). Mutations in a miRNA target site may cause an escape from miRNA repressive regulation and thus result in protein overexpression (Chin et al. 2008). Alternatively, a mutation in the 3'UTR of a gene can create a new active miRNA binding site, negatively affecting gene expression (Ramsingh et al. 2010).

During my doctoral studies, I focused my work on the formulation of the models and development of statistical methods for the discovery of tandem repeats in the eukaryotic genomes. I have obtained exact formulas for the P-values of occurrence of fuzzy tandem repeats (FTRs) and developed an FTR identification algorithm implemented in the TandemSWAN software (Boeva et al. 2006). The TandemSWAN website (http://favorov.bioinfolab.net/swan/tool.html) allows researchers to annotate exact and fuzzy tandem repeats in genomic sequences. It is usual to mask such repeats in order to avoid artifact discoveries, for example, during the analysis of next-generation sequencing data.

My post-doctoral project, which was aimed at the *in silico* discovery of TFBSs, continued my work on genomic sequence analysis. The main results are discussed in section 2.1. Section 2.2 describes my work on the methodology of TFBS detection using experimental ChIP sequencing (ChIP-seq) data. Section 2.3 focuses on the methodology of ChIP-seq data analysis for the detection of histone modifications, which we developed with a PhD student Haitham Ashoor who worked under my co-supervision.

## 2.1 *In silico* detection of transcription factor binding sites (TFBSs)

As it was stated above, transcription factor motifs are sets of DNA sequences that have higher affinity for binding by transcription factors. Further, each occurrence of a binding motif in the DNA sequence is referred to as motif hit. In the case of the direct binding of transcription factor to DNA, its binding site usually contains one or more hits of the corresponding binding motif.

There are several ways to define the set of sequences constituting a binding motif.

**Enumeration** All sequences with a potential to be bound by a transcription factor could be enumerated. The information about these sequences can be obtained from SELEX experiments (Oliphant, Brandl, and Struhl 1989). One of the drawbacks of this type of the binding motif description is that it does not allow discriminating sequences with strong and weak binding affinities.

**Consensus** An alternative model for motif description is a consensus motif constructed using the nomenclature of the International Union of Pure and Applied Chemistry (IUPAC):

```
A = adenine              C = cytosine
G = guanine              T = thymine
U = uracil               R = G A (purine)
Y = T C (pyrimidine)     K = G T (keto)
M = A C (amino)          S = G C (strong bonds)
W = A T (weak bonds)     B = G T C (all but A)
D = G A T (all but C)    H = A C T (all but G)
V = G C A (all but T)    N = A G C T (any)
```

For instance, the IUPAC consensus for the binding motif of transcription factor PU.1/Spi-1 can be written as RRVRGGAASTS (compare with the PU.1/Spi-1 logo in Figure 3) (Ridinger-Saison et al. 2012). The shortcoming of this way of modeling binding motif is that many functional binding sites may not be included in the motif.

**Position Weight Matrix** Position weigh matrix (PWM) is the most frequently used mathematical model of binding motif (Stormo 2000). PWM contains the information about the position-dependent frequency of each nucleotide in the motif. This information is represented as log odds weights $\{w_{\alpha,j}\}$ for computing a match score: $w_{\alpha,j} = log_2(F_{\alpha,j}/b_\alpha)$, where $F_{\alpha,j}$ is frequency of nucleotide $\alpha$ at position $j$ and $b_\alpha$ is the background probability nucleotide $\alpha$. Small sample correction is usually included in $F_{\alpha,j}$ to avoid taking logarithm from zero. A match score for a sequence $A = a_1 a_2 \ldots a_k$ is computed as $S_A = \sum_j w_{a_j,j}$.

PWMs can be visualized with PWM logos (Figure 3). The total height of each bin is the information content in bits of the corresponding position: $H_j = 2 - \sum_\alpha F_{\alpha,j} log_2(F_{\alpha,j})$. The height of each nucleotide is proportional to its frequency $F_{\alpha,j}$.

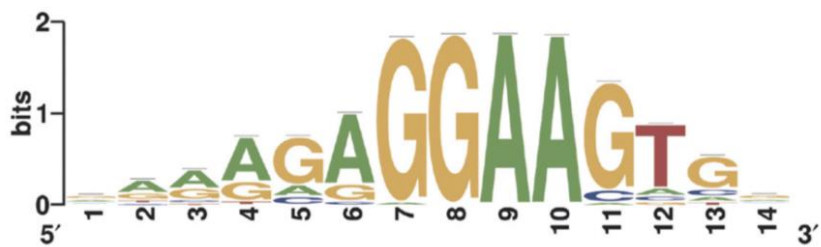

**Figure 3**. Sequence logo of the PWM created by ChIPMunk (Kulakovskiy et al. 2010) using 17,781 binding sites predicted for PU.1/Spi-1 using ChIP-seq data (Ridinger-Saison et al. 2012).

PWMs can be experimentally determined from SELEX experiments or computationally discovered from ChIP-seq data.

Using PWM representation of motif, it is possible to distinguish strong binding sites (high PWM score) and weak binding sites (moderate PWM score). It may however be a problem to distinguish weak binding sites from the background (low or negative PWM score). Usually, a cutoff on the PWM score is used to decide whether a given sequence matches the motif. The choice of this cutoff is a complex statistical task. In section 2.1.1, I describe a method we developed to define the most biologically relevant cutoff for motif PWMs.

**Bayesian networks**   Although PWM is the most widely used mathematical representation of transcription factor specificity, it still has drawbacks. PWM assumes that the positions within the binding site are independent, and the contribution at one position of the binding site to the overall affinity does not depend on the identity of nucleotides in other positions of the site. Modeling position dependencies with Bayesian networks provides an elegant solution to this problem (Barash et al. 2003; Grau et al. 2006). However, since there is no easy way to visualize motifs defined as a Bayesian network, this approach remains rarely used by the research community.

Further, we choose the PWM model to represent binding motifs. Given that a cutoff is correctly selected, we assume that transcription factor binds DNA sequences with the PWM scores higher than the cutoff.

*In silico* detection of TFBS may be separated into two tasks: detection of binding sites of transcription factors with known binding motifs (PWMs) and *de novo* motif discovery. Sections 2.1.1 and 2.1.2. focus on these two questions.

## 2.1.1. Detection of binding sites of transcription factors with known PWMs

There exist several public and commercial databases storing the PWMs for known transcription factor binding motifs. The majority of these PWMs were calculated based on ChIP-seq and SELEX experiments.

- HOCOMOCO: a comprehensive collection of human transcription factor binding sites models (Kulakovskiy et al. 2013).
- JASPAR 2014: an extensively expanded and updated open-access database of transcription factor binding profiles (Mathelier et al. 2014).
- SwissRegulon: a database of genome-wide annotations of regulatory sites (Pachkov et al. 2007).
- TRANSFAC®: a commercial database on TFBSs, PWMs and regulated genes in eukaryotes (Matys et al. 2006).

True binding sites usually score high with the corresponding PWM, while the background sequences have low PWM scores. It is not sufficient to scan a DNA region to get a PWM score at each position. The main difficulty is to correctly set the cutoff on the PWM score to separate the true binding sites from the background. Evaluation of statistical significance of motif hits can help to solve this issue (Boeva et al. 2007).

When PWM score cutoff $c$ is given, it is possible to enumerate all possible sequences matching PWM with a score above the cutoff. Let us call this set $M_c = \{A_{s_1}, A_{s_2}, \ldots, A_{s_k}\}_{s_i > c}$, where $A_{s_i}$ are all nucleotide sequences with PWM score $s_i > c$. The higher the cutoff $c$, the smaller the set of motif sequences $M_c$. Given a set of regulatory regions (enhancers or promoters) $R$, we can define the number $N_{R,c}$ showing how many $A_{s_i}$ from $M_c$ occurred in $R$. With a higher cutoff, fewer motif hits will be detected; the corresponding binding sites are likely to have strong binding affinity. With a lower cutoff, more motif hits are detected; they would correspond to both strong and weak binding sites.

In regulatory regions, binding sites often tend to occur in clusters and are overrepresented in the set $R$ of regulatory sequences targeted by the transcription factor. It is not the case for random sequences. The procedure we developed to specify the cutoff on PWM score for set $R$ is based on this assumption.

The significance of motif hit overrepresentation can be measured through the P-value, i.e. the probability to observe at least the same number $N_{R,c}$ of motif hits with cutoff $c$ in a random sequence of the total length equal to the total length of sequences in $R$ (Figure 4). Setting different cutoffs $c$, one would get different number of motif hits $N_{R,c}$ in $R$ and different P-values, $P(M_c, N_{R,c})$. Minimum of $P(M_c, N_{R,c})$ over $c$ will provide a cutoff corresponding to the most significant motif overrepresentation in R. This approach can be equally applied to several PWM corresponding to several transcription factor binding motifs (Figure 5).

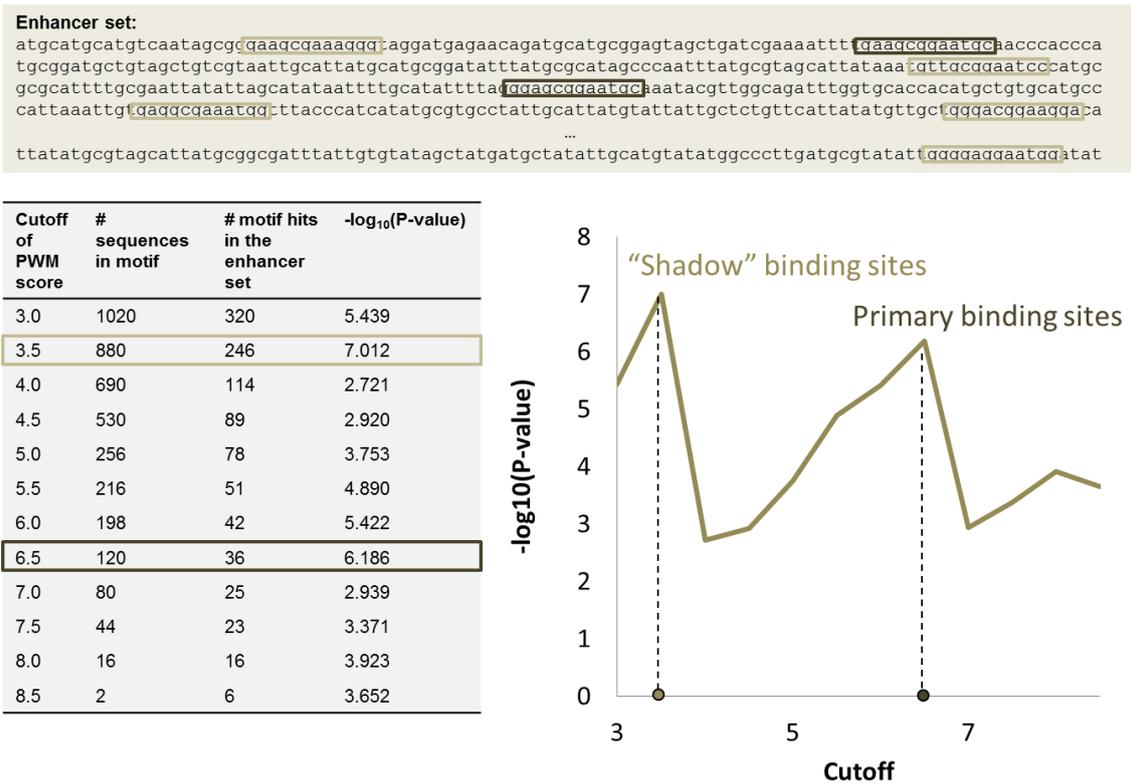

**Figure 4**. PWM score cutoff selection for a set of enhancer regions. Two local maxima in the P-value graph provide two P-value cutoffs that correspond to primary binding sites (high cutoff) and "shadow" binding sites (low cutoff). The table shows how many potential k-mer sequences match the PWM with a given cutoff (column 2), the number of motif hits in the set of enhancers (column 3) and the corresponding P-value (column 4).

The exact P-value calculation for multiple motifs with overlapping (and self-overlapping) motifs is a difficult computational task. The compound Poisson distribution formula for the P-value provides a good approximation, but not in the case of several highly overlapping motifs. For overlapping motifs, one needs a more precise method for the P-value calculation. We developed an exact algorithm for P-value calculation for the general case of heterotypic clusters of motifs, which was based on the Aho-Corasick automaton and employed a prefix tree together with a transition function (Boeva et al. 2007). This statistical approach was implemented in C++ software: Aho-Corasick Probability calculator (AhoPro, http://favorov.bioinfolab.net/ahokocc/). AhoPro enables detection of transcription factor binding sites in a given DNA sequence and calculation of statistical significance of the results. It accepts six models for transcription factor binding sites and can calculate P-values for several (possibly overlapping) transcription factor binding motifs. Zero or first order Markov chains can be adopted as null models for the random text.

This approach was applied to the identification of binding sites of cooperatively and anti-cooperatively functioning regulatory proteins in *D. melanogaster* (Boeva et al. 2007). By employing this method, we discovered the phenomenon of "shadow" TFBS in the enhancers of the *D. melanogaster* genome. Shadow binding sites are low affinity binding sites that alone are not capable to retain the transcription factor long enough to ensure the activation/repression function but instead are used to maintain a high concentration of transcription factor in the vicinity of the primary binding sites.

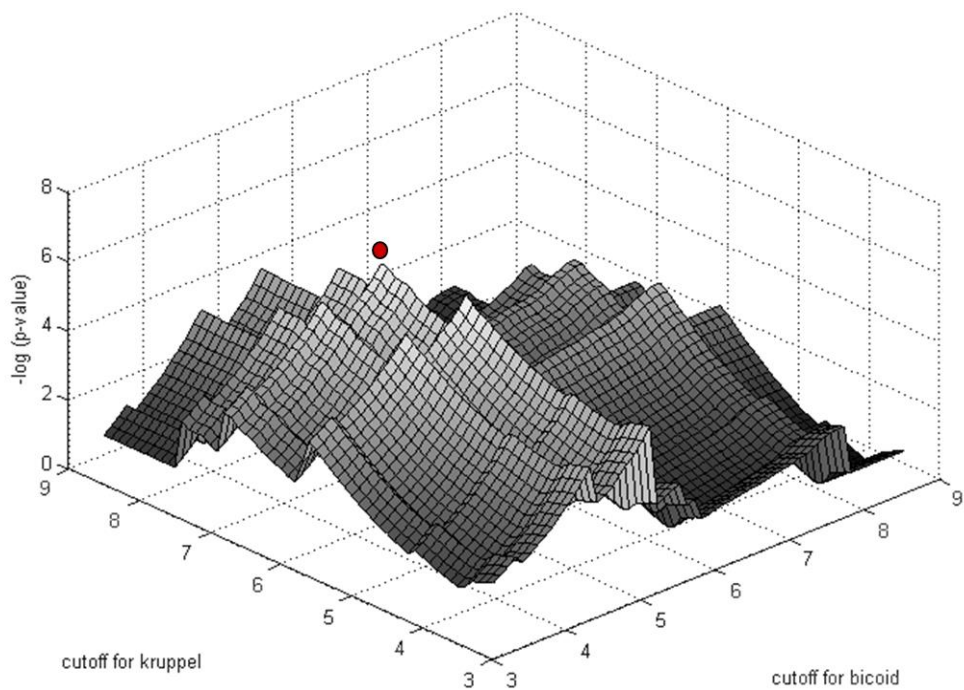

**Figure 5**. Simultaneous PWM score cutoff selection for PWMs of two *D. melanogaster* transcription factors: Bicoid and Krüppel. The graph shows the distribution of $\log_{10}$(P-value) as a function of the cutoff for the two PWMs for the enhancer of the gene even-skipped stripe 2 (*eve2*). The red point corresponds to the most significant combination of PWM and cutoffs.

We should mention that the choice of the background model is very important. Using Markov chains allows capturing the dependencies between nucleotides in the background. For instance, one can take into account the low or high frequencies of CpG nucleotides in the set of enhancer or promoter sequences.

### 2.1.2. *De novo* motif discovery

When the PWM of the transcription factor of interest is not known, it can be obtained using *de novo* motif discovery from a set of DNA sequences containing binding sites of this transcription factor. The technique consists of defining the most overrepresented motif in a given set of DNA sequences. The set of DNA sequences containing TFBSs of a particular protein can be obtained with SELEX or ChIP-x experiments (i.e. ChIP-seq, ChIP-exo, ChIP-on-chip). ChIP-Seq (Johnson et al. 2007) and ChIP-exo (Rhee and Pugh 2011) consist of the immunoprecipitation of protein–DNA complexes followed by massively parallel sequencing of short ends of immunoprecipitated DNA. These techniques succeeded the ChIP-on-chip technique and have nearly replaced the latter because of the increased accuracy in identification of TFBSs. Below I will focus on the *de novo* motif discovery for the ChIP-seq technique.

ChIP-seq yields a set of DNA regions (also called peaks) where each position has a weight reflecting how often a given DNA region was cross-linked with the protein of interest during ChIP stage (coverage profiles). The output of a ChIP-seq experiment can include tens of thousands of peaks, some longer than 1000 bp.

For *de novo* motif detection, we designed ChIPMunk, an iterative algorithm that combines greedy optimization with bootstrapping and uses coverage profiles as motif positional preferences (Kulakovskiy et al. 2010). ChIPMunk does not require truncation of long DNA segments and it is practical for processing up to tens of thousands of data sequences. The comparison with traditional (MEME) (Bailey et al. 2009) or ChIP-Seq-oriented (HMS) (Hu et al. 2010) motif discovery tools showed that ChIPMunk identified the correct motifs with the same or better quality than other tools but worked dramatically faster. Also, Ma *et al.* provided an extensive comparison of *de novo* motif discovery tools capable of using ChIP-seq data, where they demonstrated that ChIPMunk and POSMO were the best-performing tools (Ma et al. 2012).

Usually, we are interested in the discovery of *several* binding motifs. It can be several binding motifs for the same transcription factor or co-factor binding motifs. For these two cases, different motif discovery procedures should be applied. These two procedures are implemented in ChIPMunk as "Mask sequences" and "Mask motifs" modes. The first motif identified is always the motif with the highest Kullback discrete information content (KDIC). Then, the second motif is identified as the motif with the highest KDIC either in the sequences that do not contain the first motif ("Mask sequences" mode), or in the total set of sequences where the hits of the first motif have been masked ("Mask motifs" mode) (Figure 6).

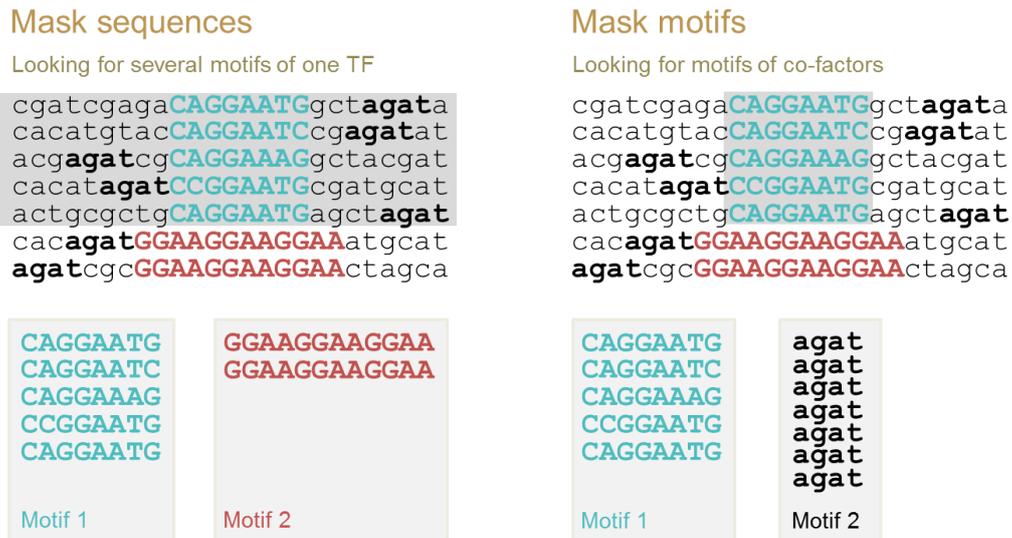

**Figure 6**. Two modes of multiple motif detection: "Mask sequences" mode to discover binding motifs of the same transcription factor, and "Mask motifs" mode to discover binding motifs of co-factors. After the first motif is identified, either all sequences containing this motif hits are removed from further analysis (sequences in grey, "Mask sequences" mode), or motif hits are masked (motif hits in grey, "Mask motifs" mode). The second motif is defined as the motif with the highest KDIC in the remaining nucleotide sequences.

The underlying assumption to use the "Mask sequences" mode is that the same transcription factor can, in some cases, bind to significantly different binding motifs; but almost each binding site should contain at least one instance of one of the binding motifs (Jie Wang et al. 2012). We should mention that frequently a transcription factor has only one binding motif; the higher the PWM score of the corresponding motif, the stronger the binding affinity (Kulakovskiy et al. 2013; Kulakovskiy et al. 2010). In this case, the "Mask sequences" mode will output only one motif. This motif will be present in almost all sequences from the set. The situation when the same transcription factor has different binding motifs can also occur. For instance, it is the case of EWS-FLI1 (Guillon et al. 2009) or NRSF (Johnson et al. 2007) (Figure 7). Also, some proteins, such as PU.1, can bind to DNA both directly and indirectly (Figure 2). In these cases, the "Mask sequences" mode will provide as a result several motifs; it can be the motifs for the direct and indirect binding (e.g. motifs for PU.1 and GATA1 for the situation illustrated in Figure 2).

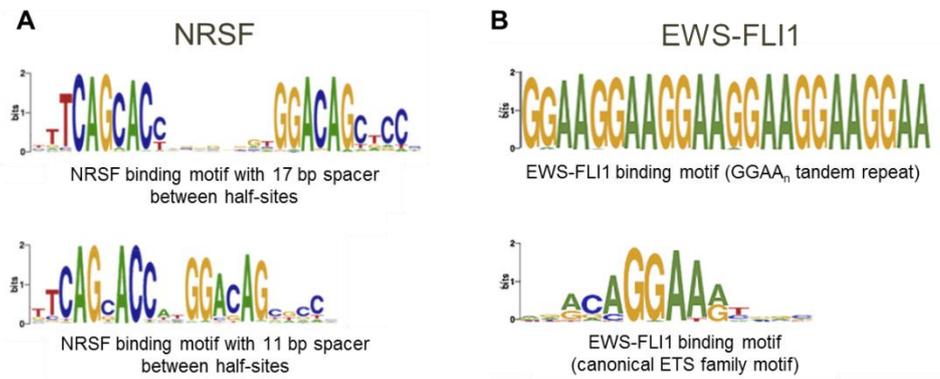

**Figure 7**. Transcription factor can have several binding motifs. (**A**) Logos for binding motifs of transcription factor NRSF with 11 bp and 17 bp spacer between half-sites (Johnson et al. 2007); (**B**) Logos for binding motifs of chimeric transcription factor EWS-FLI1 (Guillon et al. 2009; Boeva et al. 2010).

The underlying assumption for the utilization of the "Mask sequences" mode is that co-factors of the main transcription factor bind close to the main transcription factor in the regions detected with the chromatin immunoprecipitation of the main transcription factor (Figure 6, right panel). Thus, binding motifs of co-factors can be detected as over-represented motifs after the motif hits of the main transcription factor have been masked.

When binding motif is identified *de novo*, it is possible to compare its PWM or IUPAC consensus with the known motif PWMs stored in the transcription factor motif databases (JASPAR: http://jaspar.genereg.net/, Motif Comparison Tool of the MEME Suite: http://meme.nbcr.net/meme/cgi-bin/tomtom.cgi).

## 2.2 *In vivo* detection of transcription factor binding sites

In the previous sections, I described the methods for motif discovery when a set of enhancer or promoter sequences containing the binding motif was given. However, the definition of this dataset is per se a challenging topic. The most common way to define sequences containing transcription factor binding sites is to perform ChIP-seq experiments and call regions enriched in the ChIP-seq signal.

At the completion of a ChIP-Seq experiment, millions of short (∼35–75 bp) directional DNA reads are obtained; they can be positioned (or aligned) to the reference genome of the sample organism (Johnson et al. 2007). Each short read represents an extremity of a longer DNA fragment (usually, ∼200–400 bp) isolated from the chromatin immunoprecipitation. The fragment length distribution is determined by a DNA shearing technique, usually sonication, preceding chromatin immunoprecipitation. Most of the reads, when extended to the length of the initial DNA fragments, were bound by the protein of interest. By extending each read, it is then possible to identify areas of overlap of immunoprecipitated fragments and locate the protein-DNA binding event. This is usually done through the calculation of the density profile of DNA fragment coverage. Regions with high density are called peaks. The procedure of peak calling through read extension (Figure 8A) was first implemented in the FindPeaks software (Fejes et al. 2008).

Another way to calculate density profiles is implemented in one of the most used peak calling software, MACS (Zhang et al. 2008). MACS evaluates the distance between clusters of forward and reversed mapped reads, then shifts the reads by half of this distance and calculates the density as the number of overlapping shifted reads per position (Figure 8B). This procedure is the first step of the peak calling through read shift.

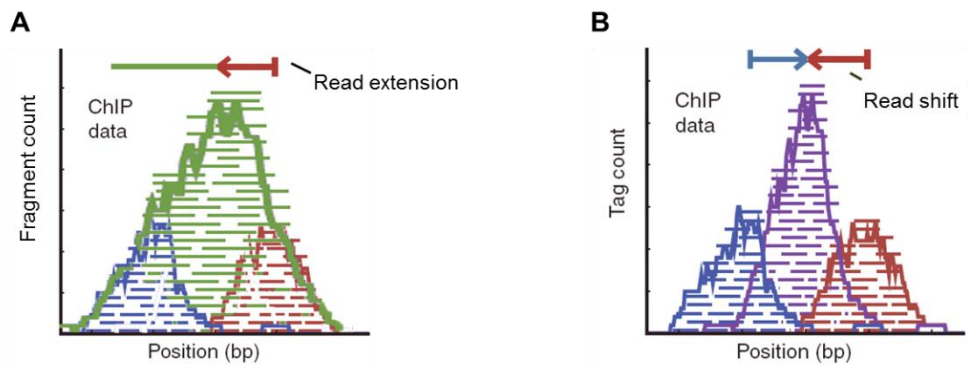

**Figure 8**. Calculation of density profile using read extension (**A**) and read shift (**B**, adopted from (Pepke, Wold, and Mortazavi 2009)).

Not every peak (region enriched in overlapping extended or shifted reads) contains a true binding site. Low peaks (with moderate read density) can occur by chance. Thus, to characterize the read enrichment and discriminate true binding from background noise, a statistical model should be applied. To this moment, there are more than 20 different tools that perform this task for ChIP-seq data for transcription factors (Wilbanks and Facciotti 2010; H. Kim et al. 2011). The background model can be based on the uniform distribution of sequenced reads along the genome. Under such a background model, the Poisson P-value can be applied to evaluate the significance of read over-representation in a given region (Zhang et al. 2008). Often a negative control experiment is performed to assess the distribution of sequenced reads in the background. The last studies showed that an appropriate control data set is critical for analysis of any ChIP-seq experiment because DNA breakage during sonication is not uniform (Landt et al. 2012). Two types of control experiments are commonly used. The first one is sequencing "Input DNA", i.e. perform all steps of ChIP-seq library preparation including cross-linking and fragmentation without chromatin immunoprecipitation. The second one is performing a ChIP-seq experiment using a nonspecific immunoglobulin G (IgG) antibody ("IgG" control).

In addition to ununiformed DNA breakage during the DNA shearing stage, other factors can contribute to the clustering of background reads in the control and ChIP experiments. The most prominent one is misalignments of reads coming from repetitive regions. Some tools provide an explicit option to exclude repetitive regions known to contain clusters of reads from the background (Nix, Courdy, and Boucher 2008; Boeva et al. 2010). Amplification of genomic DNA (e.g. due to copy number aberrations in cancer) results in high read density in both the control and ChIP datasets in the corresponding regions. Thus, one needs to formulate an appropriate statistical model for the detection of binding sites in ChIP-seq data to take the above biases into consideration.

In 2010, we demonstrated that the accuracy of peak calling can be considerably improved by incorporating information about genomic sequences of peaks in addition to coverage depth information (Boeva et al. 2010). We presented an algorithm based on the idea that functional binding sites of transcription factors should contain a consensus motif (or a set of consensus motifs). The algorithm was implemented in the MICSA software (Motif Identification for ChIP-Seq Analysis).

The MISCA workflow consists of four main steps: (i) identification of all candidate peaks using read extension, (ii) identification of binding motif PWMs from a subset of peaks, (iii) detection of motif hits in all candidate peaks, and (iv) optimization of the peak calling output by calculating statistics taking into account information about both motif hit and depth of coverage.

The set of candidate peaks obtained after the first step contains a large number of false positive predictions. To filter them out, other methods commonly determine a threshold on DNA read density (Valouev et al. 2008; Ji et al. 2008; Kharchenko, Tolstorukov, and Park 2008). However, our experience and that of other researchers showed that even low candidate peaks (i.e. regions with low read density) often contain functional binding sites. We proposed an approach based on the presence of motif information in peak sequences. We used this information to retain additional peaks containing strong motifs, especially those with low read density. For each subset of peaks with a given read density, the algorithm chooses a criterion for peak retention based on motif strength and position within the peak. Then, the number of false positive peaks expected to satisfy the criterion is evaluated. The algorithm contains an optimization procedure applying different criteria in an attempt to maximize the total number of selected peaks for all subsets without having the estimated total number of false positive peaks exceed the user defined threshold.

MICSA is able to automatically identify overrepresented TF binding motifs. Importantly, MICSA identifies *several* binding motifs, which, as we showed with the EWS-FLI1 example, can possibly carry different biological functions. For detect several PWMs of motifs overrepresented in the set of candidate ChIP-seq peaks, MICSA uses the "Mask sequences" mode described in section 2.1.2.

The statistics calculated by MICSA allows keeping high peaks (i.e. regions with high number of immunoprecipitated fragments) as well as low and moderate peaks with strong motif hit in the peak center (Figure 9) in the final peak set. Low peaks without strong motif hit will be removed from the final dataset.

Since MICSA checks for motif occurrences in all peaks including those with very low coverage depth, there is no need in the explicit selection of threshold on read density. The only parameter that remains to be specified is the maximal number of expected false positive hits among selected peaks or the maximal false discovery rate (FDR).

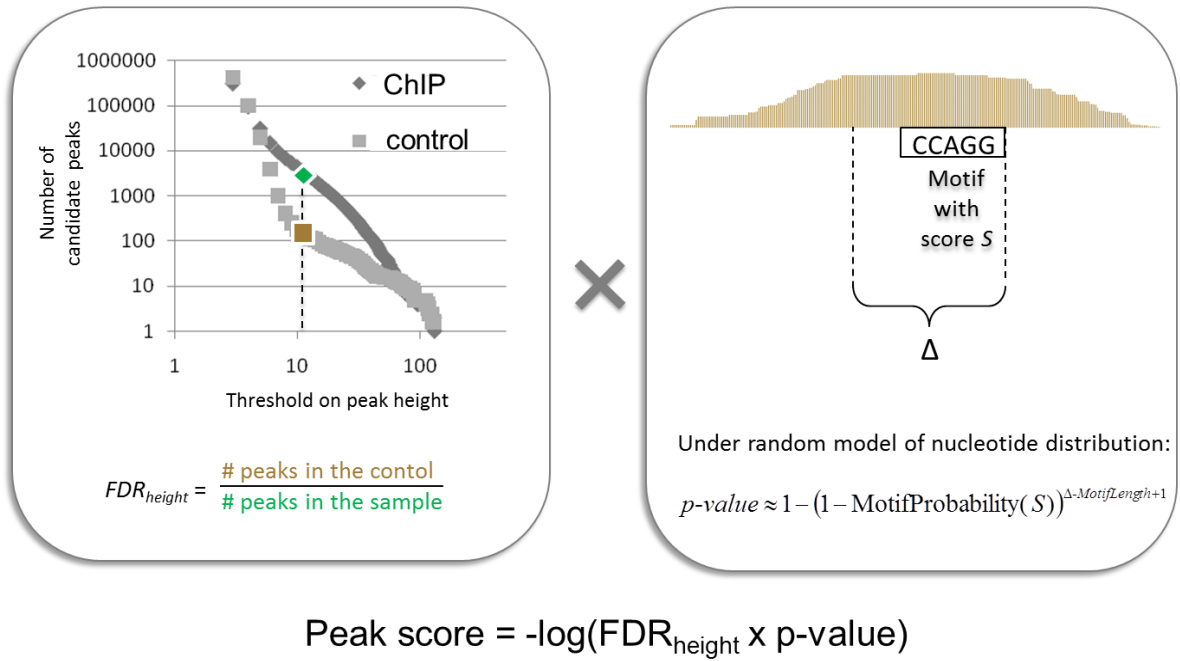

Peak score = -log(FDR$_{height}$ x p-value)

**Figure 9**. Illustration of the procedure of peak score calculation employed by the MICSA algorithm.

In the corresponding paper (Boeva et al. 2010), we demonstrated that MICSA provided better results in the terms of sensitivity of binding site detection compared to ten other peak calling tools: MACS (Zhang et al. 2008), PeakSeq (Rozowsky et al. 2009), QuEST (Valouev et al. 2008), spp/wdt (Kharchenko, Tolstorukov, and Park 2008), Useq (Nix, Courdy, and Boucher 2008), F-Seq (Boyle et al. 2008), CisGenome (Ji et al. 2008), ERANGE/ChIPSeqMini (Mortazavi et al. 2008), SISSRs (Jothi et al. 2008) and FindPeaks (Fejes et al. 2008). We also demonstrated the high reproducibility of peaks called by MICSA.

The main steps of the most typical ChIP-seq analysis were assembled in a web-based tool, Nebula (Boeva, Lermine, et al. 2012). Nebula is based on the Galaxy open source framework. Galaxy already includes a large number of functionalities for mapping reads and peak calling. We added the following to Galaxy: (i) peak calling with FindPeaks and a module for immunoprecipitation quality control, (ii) *de novo* motif discovery with ChIPMunk, (iii) calculation of the density and the cumulative distribution of peak locations relative to gene transcription start sites, (iv) annotation of peaks with genomic features and (v) annotation of genes with peak information. Nebula generates the graphs and the enrichment statistics at each step of the process. During steps iii–iv, Nebula optionally repeats the analysis on a control dataset and compares these results with those from the main dataset.

MICSA and new functionalities including the possibility of analysis of cancer ChIP-seq datasets are going to be added to Nebula during the internship in our lab of a master student Amira Kramdi, who will work under my supervision in summer 2014.

### 2.2.1. Transcriptional regulation by oncogenic EWS-FLI1 in Ewing sarcoma

An example of ChIP-seq data analysis using the methods described in the previous sections is provided by the study of mechanisms of transcriptional regulation by EWS-FLI1, a chimeric protein causing Ewing sarcoma (Guillon et al. 2009; Boeva et al. 2010).

Ewing tumors, the second most frequent bone tumors in teenagers and young adults, show specific translocations fusing the 5' part of EWS to the 3' sequence encoding the DNA binding domain of an ETS factor (Arvand and Denny 2001). In most cases, translocations occur between chromosomes 11 and 22, leading to the formation of the aberrant EWS-FLI1 chimeric transcription factor (Delattre et al. 1992). In rarer cases, ERG, E1AF, ETV1 or FEV that encode other ETS family members are fused to EWS. The chimeric transcription factor EWS-FLI1 has, as a result of a fusion of two genes, DNA binding domain of transcription factor FLI1 and activation domain of RNA binding protein EWS. EWS-FLI1 can recognize *in vitro* the same sequences as FLI-1 (Mao et al. 1994), but is a more potent transactivator than the wild type factor. This abnormal transcription factor is a key oncogene in Ewing sarcoma.

In an attempt to decipher a general EWS-FLI1 DNA binding mechanism and to identify candidate direct target genes in the Ewing tumor context, we have combined ChIP-Seq for EWS-FLI1 and the analysis of EWS-FLI1-induced gene expression modulation (HG133A Affymetrix gene expression array). For the ChIP-seq, we used a FLI1-specific antibody that binds both FLI-1 and EWS-FLI1. Since FLI-1 is not expressed in Ewing cells, we did not expect off-target effects. A control was obtained using the same anti-FLI1 antibody in a rhabdoid tumor cell line (MON) that does not express EWS-FLI1 or ETS family TFs.

In our case, the total amount of sequenced DNA was insufficient for straight-forward identification of the majority of binding sites. Peak calling using the FindPeaks software (Fejes et al. 2008) showed a very limited number of regions of EWS-FLI1 specific binding (246 binding sites); the assessment of the FDR using the background model based on the peak height distribution in the control dataset resulted in evaluated FDR of 20% for this set of 246 binding sites. Upon the analysis with the MICSA algorithm, we were able to discover 2,264 potential binding sites with FDR evaluated at the level of 5% using the background model that takes into account peak height distribution in the ChIP and control datasets and genomic nucleotide content.

MICSA was able to identify two consensus motifs occurring in the DNA sequences of peaks called (Figure 7B). The first motif represented a (GGAA)$_{\geq 6}$ microsatellite and it was found in 496 peak sequences; the second motif, found in 1,768 peak sequences, corresponded to the consensus RCAGGAARY (R = A/G, Y = T/C). Since the single RCAGGAARY consensus motif resembles the known binding motifs for the ETS TF family, we refer to it as the ETS motif. Interestingly, the extracted ETS motif, although resembling the consensus motif of FLI1 (CCGGAARY), did not completely coincide with the latter, even though EWS-FLI1 shares the same DNA-binding domain as FLI1.

We compared the locations of peaks discovered with gene expression data for the Ewing cancer cell line A673 in both the presence and the absence of EWS-FLI1 using a random set of peaks as a control (Figure 10). To create the random set (grey in Figure 10) we randomly selected 2,264 locations in the annotated part of the human genome (NCBIv36, hg18). From the expression data, we extracted a list of putative target genes of EWS-FLI1: 557 genes down-regulated by EWS-FLI1 and 577 up-regulated genes (fold change >|2| with a Welsh P-value <0.01).

Our analysis revealed the tendency of sites bearing GGAA-microsatellites to activate the expression of neighboring genes [sites found from 150-kb upstream to 50-kb downstream of gene transcription start sites (TSSs)] (Figure 10A), while sites with the ETS motif do not seem to have a definite activator function (Figure 10B). ETS sites show some transcriptional inhibitory influences on gene expression when located in the first 50-kb downstream of the TSSs. However, when ETS-sites are found further away from genes (within 1 Mb upstream or downstream but not in the first 50-kb downstream TSS), both activatory and inhibitory influences are observed for EWS-FLI1 transcriptional activity. Among other hypotheses, this could be explained by competitive binding of EWS-FLI1 and native repressor or activator transcription factors.

Our analysis confirmed binding sites for five known direct target genes of EWS-FLI1: *MYC*, *CCND1*, *TGFBR2* (Fukuma et al. 2003), *CAV1* (Tirado et al. 2006) and *IGF1* (Cironi et al. 2008). In two cases out of five, a binding site was identified inside the gene and not in the promoter region. MICSA predicted many new possible direct targets of EWS-FLI1. Among them we found *PPP1R1A*, *LBH*, *FAS*, *CAV2* and *NBL1*. This information helped to construct a detailed regulation network for Ewing sarcoma (Stoll et al. 2013).

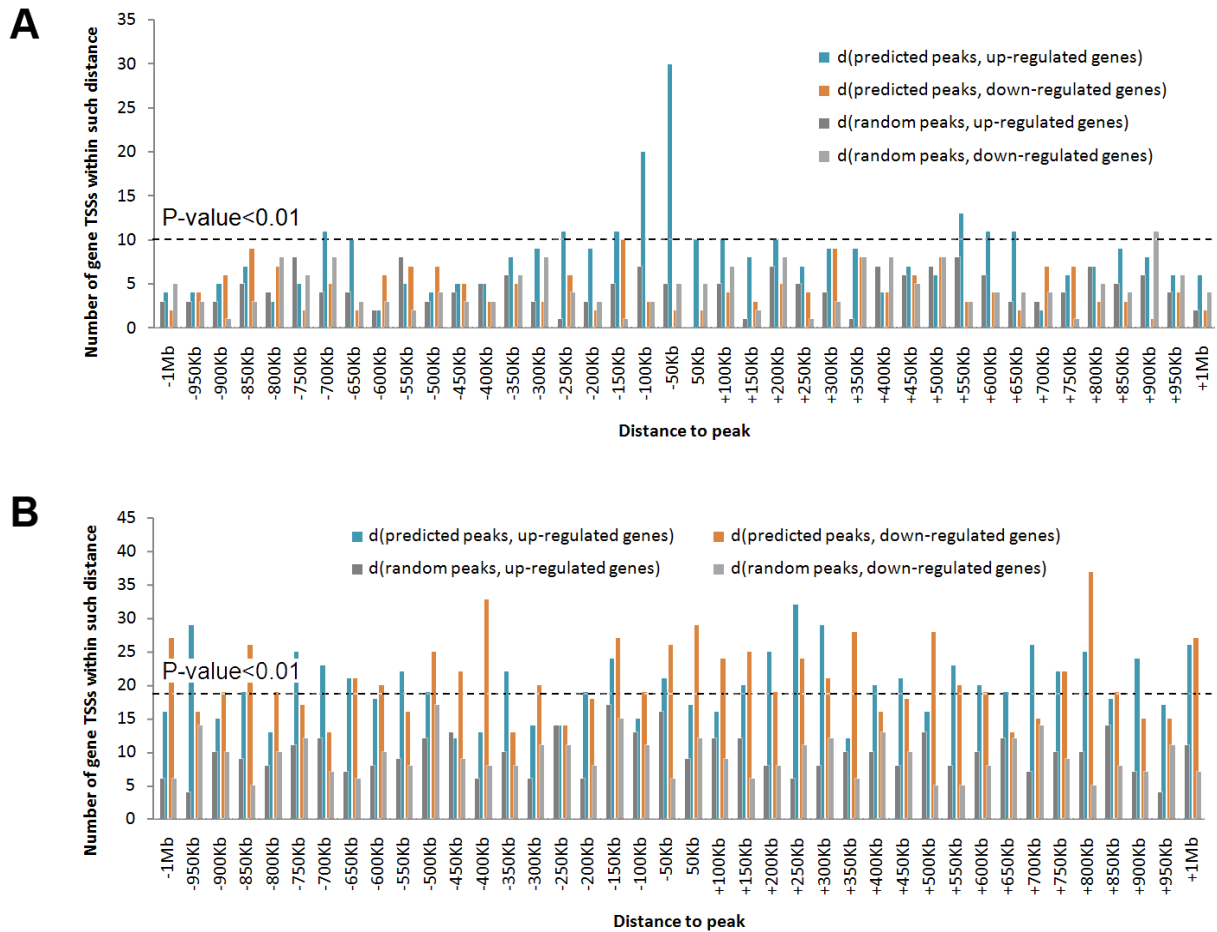

**Figure 10**. Histogram of distances between predicted/random peaks and genes up-/down- regulated by EWS-FLI1. (**A**) Predicted sites containing $(GGAA)_n$ microsatellites; (**B**) ETS sites (site without $(GGAA)_n$ microsatellites). EWS-FLI1 binding to GGAA microsatellites results in significant expression activation of neighboring genes. EWS-FLI1 binding to single ETS sites can produce both negative and positive effects on transcription of neighboring genes. The *P*-values were directly evaluated by Monte-Carlo simulations of random peaks. Distances from the TSSs of modulated genes to random peaks (iterative trials) and to predicted sites were calculated. The *P*-values correspond to the probability to get at least the observed number of distances falling within a given 50-kb window, under the hypothesis that peaks are randomly distributed and their coordinates are independent of coordinates of TSSs of EWS-FLI1 modulated genes. Bars above the dashed line correspond to a *P*-value <0.01.

### 2.2.2. Transcriptional regulation by oncogenic Spi-1/PU.1 in erythroleukemia

The second example of analysis of ChIP-seq data is given by our study aimed at discovery of mechanisms of transcription regulation by Spi-1/PU.1, a transcription factor overexpressed in erythroleukemia (Ridinger-Saison et al. 2012).

Spi-1/PU.1 belongs to the same ETS transcription factor family as FLI1 (the DNA-binding partner of EWS in the chimera causing Ewing sarcoma). Spi-1/PU.1 is a major regulator of developmental processes that functions in hematopoietic stem cell and progenitor cell self-renewal as well as in the commitment and/or maturation of the myeloid and B-lymphoid cell lineages (Iwasaki et al. 2005). Spi-1/PU.1 is expressed at low level in erythroid progenitors and is down-regulated upon terminal differentiation (Back et al. 2004). Its expression beyond physiological expression levels promotes oncogenesis in erythroid lineage as evidenced by the development of erythroleukemia due to Friend virus (MEL) or to Spi-1/PU.1 transgenesis in mice. It have been shown that Spi-1/PU.1 induces resistance to apoptosis (Rimmelé et al. 2007) and accelerates gene elongation during replication (Rimmelé et al. 2010).

In order to understand how overexpressed Spi-1/PU.1 mechanistically controls transcription in the erythroleukemic cells, we combined ChIP-seq for Spi-1/PU.1 with gene expression profiling (Affymetrix gene expression arrays).

For *in vivo* ChIP-seq, we used freshly isolated spleen cells from SPI-1 transgenic mice that developed erythroleukemia. To assess the changes in gene expression upon the down-regulation of Spi-1/PU.1, the erythroleukemic cells were engineered to express anti-SPI-1 shRNAs in the presence of doxycycline. Cells were grown with or without doxycycline for 26 h or 44 h. The probesets corresponding to genes that were differentially expressed between the Spi-1/PU.1-overexpressing and Spi-1/PU.1-knockdown cells were determined by applying ANOVA with an adjusted P<0.01 (the Benjamini and Hochberg method).

To detect peaks corresponding to putative binding regions, we used the FindPeaks algorithm (Fejes et al. 2008) and the MICSA pipeline (Boeva et al. 2010). The total of 17,781 Spi-1/PU.1 peaks were kept for the further analysis. We performed quantitative PCR validation after the Spi-1/PU.1 ChIP assay (ChIP-qPCR) of randomly selected sequences from the ChIP-seq experiment. Of the 21 Spi-1/PU.1 binding sites tested, 20 were validated.

We annotated the peaks according to their functional location within RefSeq genes (http://www.ncbi.nlm.nih.gov/refseq/rsg/about/, release April 2011) using the Nebula peak-to-gene annotation module (Boeva, Lermine, et al. 2012). The gene location categories were: promoter (-1.5 kb upstream of the TSS), immediate downstream ('ImDown', +2 kb downstream of the TSS), enhancer (-30 kb to -1.5 kb upstream of the TSS), intragenic (+2 kb downstream of the TSS to the transcription end), 5 kb downstream ('5kbdown', +5 kb after the transcription end). For the peaks annotation, if one peak could be assigned to several overlapping isoforms of one gene, Nebula uses a hierarchy and assigns the peak to only one isoform. It uses the following priorities: promoter > immediate downstream > intragenic > enhancer > 5-kb downstream. In the cases of two overlapping genes, Nebula includes both entries.

As we expected, Spi-1/PU.1 binding turned out to be highly gene-associated. The total of 83% of Spi-1/PU.1 binding sites were located within a region spanning from -30 kb upstream of the TSS to +5 kb downstream of the TE of genes, that was statistically different from Input control peaks (68%,) ($\chi^2$ test, $P<10^{-3}$). Notably, the location of Spi-1/PU.1 DNA binding was significantly biased toward the promoter and immediate downstream subregions compared with the control peaks ($\chi^2$ test, P<$10^{-3}$).

Using ChIPMunk, we identified the main Spi-1/PU.1 binding motif underlying the 17,781 peak sequences bound by Spi-1 (Figure 3). Motif hits were found in 88% of the Spi-1/PU.1-bound regions in close proximity to the positions of peak maxima (±150 bp). We did not observe any differences in motif structure between high and low affinity binding sites, or between Spi-1/PU.1- repressed or activated genes, or between Spi-1/PU.1 binding sites in gene promoters, enhancers and other functional genomic regions. Thus, we concluded that in erythroleukemia Spi-1/PU.1 binds directly to DNA and has strong binding preferences for genes and, especially, gene promoters.

Interestingly, bound to a gene or even gene promoter, Spi-1/PU.1 rarely causes transcriptional modulation. Half of all mouse RefSeq genes contained Spi-1/PU.1 binding sites, i.e. within a –30-kb region upstream of the TSS to +5 kb downstream of the TE, and, only 8.1% (854 out of 10,560) of the Spi-1/PU.1-occupied genes were transcriptionally modulated. Thus, we intended to study what additional factors influence the gene modulation activity of Spi-1/PU.1.

In the activated genes, the Spi-1/PU.1 peaks were located closer to the TSS than in the repressed and non-modulated genes. Indeed, 60% of the activated genes contained Spi-1 peaks located within 5 kb regions around gene TSSs, whereas only 40% and 22% of repressed and non-modulated genes contained peaks within this distance around TSSs, respectively. Peaks in the promoters of activated genes were significantly higher than in the promoters of repressed and non-modulated genes (P-value < $10^{-5}$), showing that higher binding affinity is associated with transcriptional activation.

We found a correlation between the number of motif hits per peak and the Spi-1/PU.1 peak height (Figure 11). In agreement with this observation, the number of Spi-1/PU.1 motif hits in Spi-1/PU.1 ChIP-seq peaks in promoters of activated genes was significantly higher than in promoters of repressed or non-modulated genes (P-values <$10^{-6}$). Our analysis also indicated that Spi-1/PU.1 binding is favored at CG rich sequences, but the absence of CpG islands increases the potential of Spi-1/PU.1 to activate gene expression. Interestingly, the dependency between the binding affinity and transcriptional activation/repression is weak for the genes with CpG islands (P-value=0.02).

We also discovered that Spi-1/PU.1-bound enhancers of down-regulated genes tended to contain high number of Spi-1/PU.1 motif hits. We observed that Spi-1/PU.1 binding motifs are co-directed both in promoters of up-regulated genes and enhancers of down-regulated genes (Figure 12).

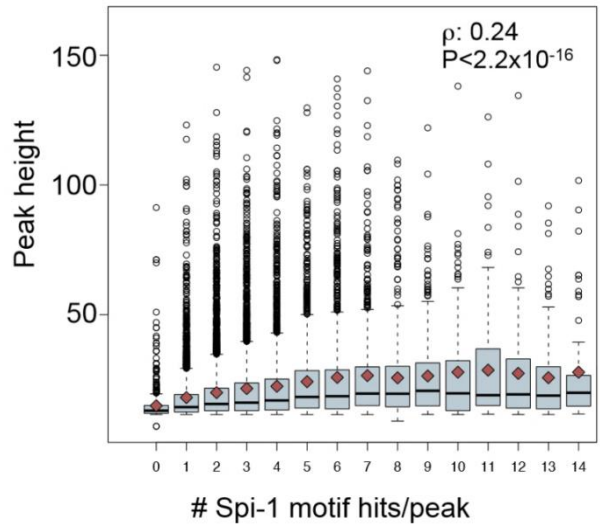

**Figure 11**. The number of Spi-1/PU.1 motif hits correlates with the Spi-1/PU.1-binding intensity measured by the peak height. The boxplot represents the distribution of the peak heights (y-axis) for each number of Spi-1/PU.1 motif hits/peak (x-axis). The dark red squares indicate the mean values, and the black line within each box indicates the median. The Spearman coefficient correlation (ρ) and the P-value of correlation test are reported.

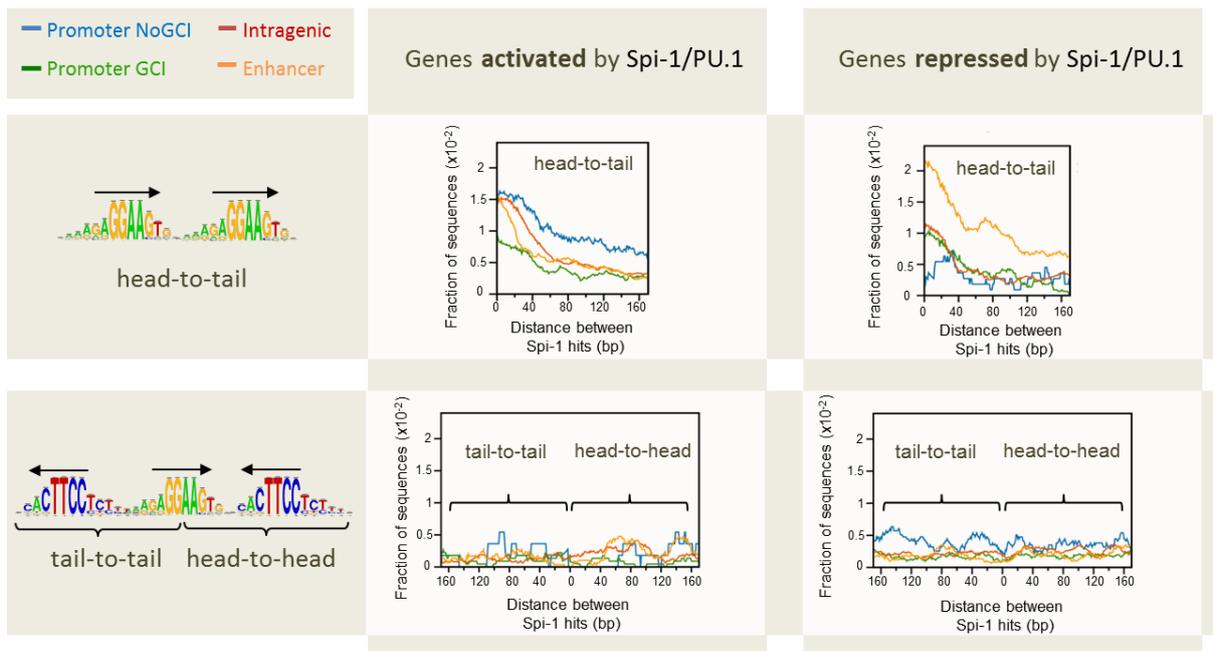

**Figure 12**. Spi-1/PU.1 motifs tend to cluster in head-to-tail orientation in the promoters of activated genes and in enhancers of repressed genes. Distribution of the distances between the pairs of Spi-1/PU.1 motif hits in head-to-tail orientation (same strand) or in head-to-head or tail-to-tail orientation (opposite strands) for activated genes and repressed genes. The X-axis presents the length of the spacers separating pairs of Spi-1 motif hits. The Y-axis shows the fraction of sequences with at least one pair of Spi-1 motif hits separated by the selected spacer. GCI (green) and NoGCI (blue) promoters specify promoter with and without CpG islands respectively.

With PATSER (Hertz and Stormo 1999), we scanned ChIP-seq peak sequences for the motif hits of known transcription factors (Transfac and Jaspar motifs libraries). We did not detect any known transcription factor motifs to be associated with differential Spi-1/PU.1 binding in promoter and enhancers, or in activated and repressed genes. However, we observed patterns in motif orientation and distance preferences between Spi-1/PU.1 motif and known binding motifs of other transcription factors including KLF family (Figure 13). The observed patterns suggest cooperative interactions between Spi-1/PU.1 and its partners. The functional significance of these observations needs to be validated by biological experiments.

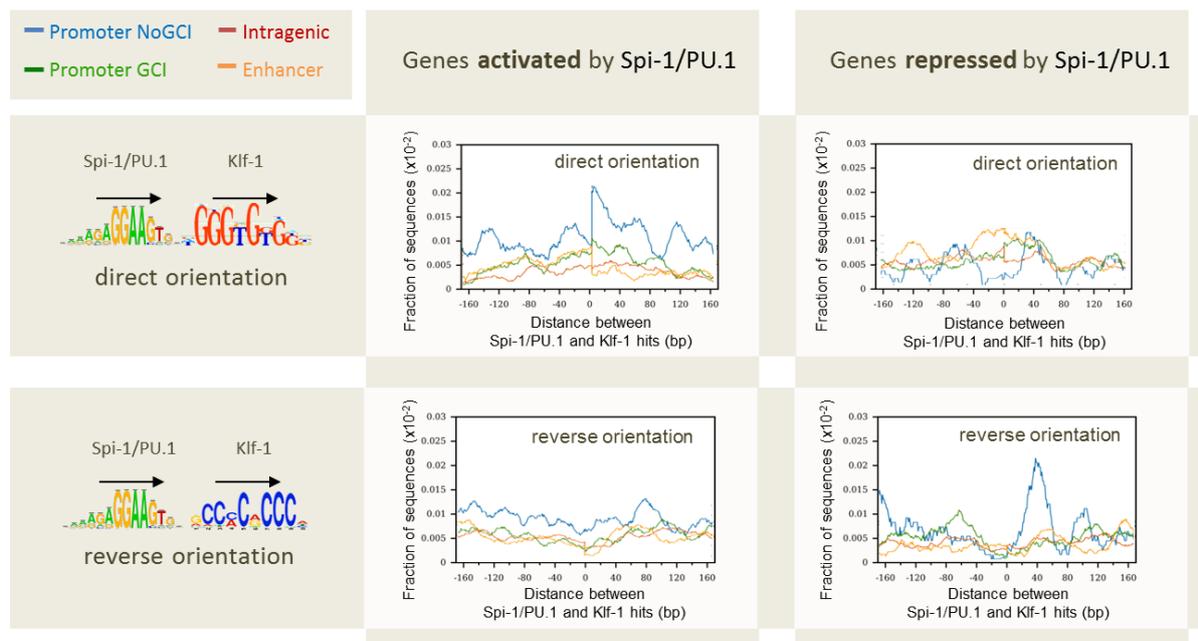

**Figure 13**. Distribution of the distances between the pairs of Spi-1/PU.1 and Klf-1 motif hits in direct or reverse orientation for genes activated and repressed by Spi-1/PU.1. The X-axis presents the length of the spacers separating pairs of Spi-1 and Klf-1 motif hits. The Y-axis shows the fraction of sequences with at least one pair of motif hits separated by the selected spacer. GCI (green) and NoGCI (blue) promoters specify promoter with and without CpG islands respectively.

Overall, our analysis identified several parameters impacting on Spi-1/PU.1-mediated transcriptional effect in erythroleukemic cells: (i) Spi-1/PU.1 binds mainly directly to DNA at a specific motif, but DNA binding does not often insinuate transcriptional activity; (ii) the transcriptional response to Spi-1/PU.1-binding is influenced by the Spi-1/PU.1 binding location in the genes; (iii) Spi-1/PU.1 occupancy at promoters of genes devoid of CGIs facilitates gene activation; (iv) tight clustering and similar orientation of Spi-1 motifs at the occupied regions favor detection of Spi-1/PU.1-mediated transcriptional activation; and (v) in contrast to the myeloid and lymphoid B cells, in the erythroleukemic cells, lineage-specific cooperating factors do not play a prevalent role in the positive control of Spi-1/PU.1-mediated transcriptional activation.

## 2.3 *In vivo* detection of histone modifications

In addition to the detection of binding sites of transcription factors, the ChIP-Seq technique is now widely used for identification of epigenetic marks such as histone variants and different covalent modifications of histone tails. Common histone modifications include lysine acetylation, methylation, ubiquitylation and sumoylation, serine and threonine phosphorylation and arginine methylation (Kouzarides 2007). As it was described in the Introduction, histone marks help partitioning the genome into euchromatin, which is accessible for transcription, and heterochromatin. For instance, trimethylation of lysine 9 of histone 3 (H3K9me3) and trimethylation of lysine 27 of histone 3 (H3K27me3) are marks associated with pericentromeric heterochromatin and regions of polycomb-mediated repression (Kharchenko et al. 2011). Also, histone modifications and histone variants are often associated with distinct biological functions. Trimethylation of lysine 36 of histone 3 (H3K36me3) is a mark of transcription elongation; trimethylation of lysine 4 of histone 3 (H3K4me3) marks active or poised promoters; monomethylation of lysine 4 of histone 3 (H3K4me1) together with acetylation of lysine 27 of histone 3 correlates with active enhancers (Kouzarides 2007). Some marks are narrow and cover 1–10 consecutive nucleosomes (e.g. H3K4me1 or H3K4me3), whereas others (e.g. H3K27me3 and H3K36me3) can cover large genomic regions, from tens to hundreds of kilobases in length. So far, except for the H3K4me3 mark, no sequence signature has been identified for histone modifications (Ha, Hong, and Li 2013).

Many tools have been developed to facilitate the analysis of histone modification data obtained with the ChIP-Seq technique. Tools designed to detect transcription factor binding sites can be applied to the detection of narrow peaks corresponding to histone modifications of type of H3K4me3 (Kharchenko, Tolstorukov, and Park 2008; Rozowsky et al. 2009; Zhang et al. 2008). Other methods are able to identify epigenetic marks covering large genomic regions. This is mostly done through clustering of narrow regions with significant read enrichment (Zang et al. 2009), gene-by-gene quantification of enrichment (Hebenstreit et al. 2011), Hidden Markov Models (HMMs) (Qin et al. 2010; H. Xu et al. 2008) and linear signal–noise models (H. Xu et al. 2010).

The methods mentioned above were designed to work with ChIP-seq data generated for normal cells. The normal cells do not contain genomic alterations inherent to cancer cells. However, it is extremely important to be able to identify histone modifications in cancer. Indeed, although genetic modifications remain the main cause of cancerogenesis, epigenetic modifications such as changes in histone modification landscape and, especially, LRES, play an important role in cancer development and progression (Esteller 2007; Stransky et al. 2006).

Genomic rearrangements frequently happening in cancer cells result in large genomic losses and gains. A DNA region present in a cell in four copies instead of two will produce twice more reads than a similar region present in only two copies; this is true both for background regions and regions enriched in the signal (histone modifications). Through the appropriate use of the control sample during the design of a statistical model for the detection of histone modifications, this fact should be taken into account.

The work accomplished by PhD student Haitham Ashoor under my supervision concerned the development of a computational method, HMCan, for the detection of regions enriched in histone modifications from cancer ChIP-seq data. The algorithm we designed is able to correctly predict histone modifications regardless of the presence of copy number alterations in the tumor sample genome (Ashoor et al. 2013).

We wished to design an algorithm able to detect epigenetic marks covering both large and narrow genomic regions (e.g. H327me3 and H3K4me4). HMM (Baum and Petrie 1966; Baum et al. 1970) was a technique that could provide an effective solution. As it was shown by Qin *et al*. (2010), HMMs can potentially capture ChIP signal both for narrow and large epigenetic marks. Through the assessment of hidden states, it possible to partition the genome into regions of ChIP signal and background. In the case of the two-state hidden model we employed, one hidden state corresponded to the enrichment in immunoprecipitated reads and the other one corresponded to the background.

The raw read density is calculated using the read extension procedure described in beginning of Section 2.2 (Figure 8). In addition to the presence or absence of ChIP signal, the raw read density values depend on many factors. I have already mentioned copy number changes to be among such factors. The higher the number of copies, the higher the observed read count. Other parameters to take into account are the read mappability and GC-content bias. These biases together with the corrections we applied are described in the next paragraphs.

To estimate the copy number of each genomic region, we use the algorithm implemented in the Control-FREEC software (Boeva et al. 2011; Boeva, Popova, et al. 2012) to the read counts in the control dataset (usually, input DNA). The sequencing of input DNA represents whole genome sequencing (WGS) and thus the WGS approaches such as Control-FREEC can be applied. The algorithm segments the genome according to the copy numbers and assigns a "ratio" value to each segment. The ratio shows to what extent the local copy number status is higher or lower than the neutral level. When ratio $r$ of each position is estimated, each value in the raw density profile calculated for the input and ChIP datasets is divided by $r$. This brings low (or high) reads counts of low- (high-) copy number regions to the same level as the read counts in the neutral region.

Sequencing technologies often result in association between number of reads mapped to a specific DNA region and the percentage of nucleotides G and C (its GC-content) (Dohm et al. 2008). In sequencing data analysis, including ChIP-seq data analysis, we usually keep only uniquely mapped reads, i.e. reads that can be positioned to the unique genomic coordinates by the read mapping algorithms. The mappability depends on the read length: the longer the read, the higher the chance that there will be just one subsequence in the genome that would match the read. Most tandem repeats in the genome are not uniquely mappable with short reads. Interspersed repeats (e.g. SINEs and LINEs) have on average lower mappability than regions not annotated as repeats in the RepeatMasker database. The authors of (Benjamini and Speed 2012) suggested a universal way to correct read counts for the GC-content and mappability. In our method, we applied this correction to remove GC-content and mappability bias, which otherwise may result in aberrant read counts.

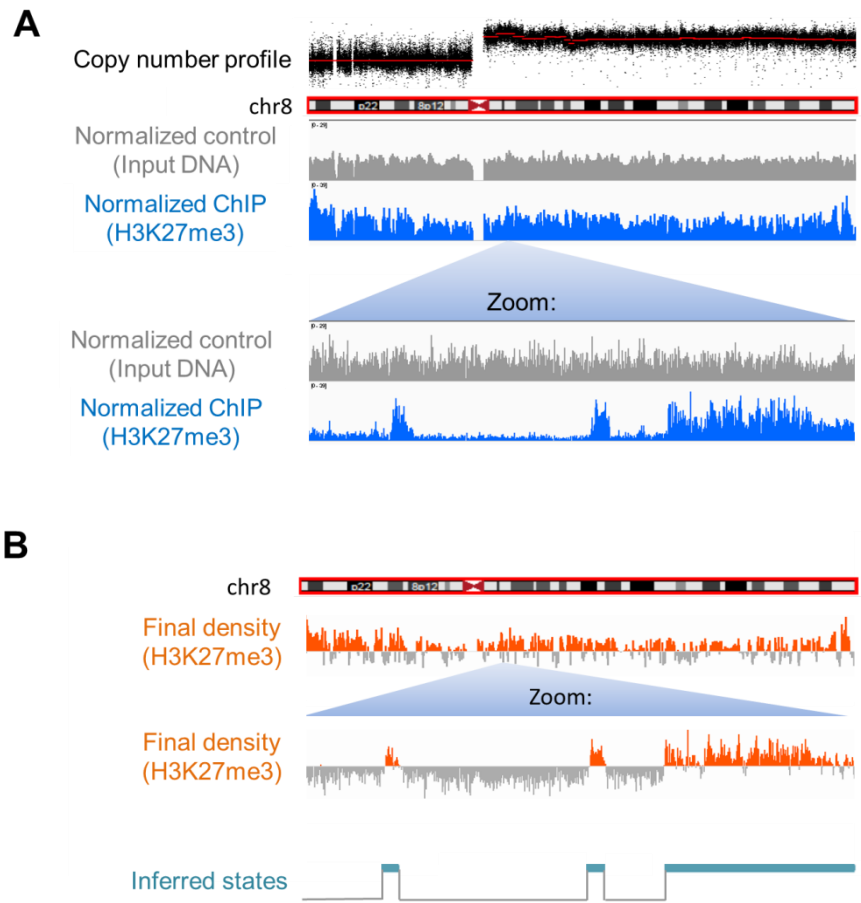

**Figure 14**. Example of the distribution of the normalized read count densities along the genome. ChIP-seq H3K27me3 data for the CL1207 human bladder transitional cell carcinoma cell line from (Ashoor et al. 2013). **A**. Top track: copy number profile for chromosome 8 calculated by GAP (Popova et al. 2009) using SNP array technology; Bottom tracks: ChIP (grey) and input DNA (blue) read count densities. The normalized densities do not show copy number bias. **B**. The result of noise subtraction from the normalized ChIP read count densities (final density). The result of two-state HMM applied to the final density profile (sky-blue).

The profiles resulting from the normalization by HMCan of the ChIP and control datasets do not have GC-content, mappability or copy number bias (Figure 14A). In order to remove other possible artifacts of sequencing and mapping we subtract the control density profile (grey in Figure 14A) from the ChIP profile (blue in Figure 14A) with a coefficient that represents the noise in the ChIP data. This coefficient is the average value of ratio between normalized densities in the ChIP and control in the putative background regions. The initial guess of signal and background regions is done by applying the Poisson test on the ChIP and control densities. The same regions are used to calculate the initial emission and transition probabilities for the HMM.

To infer the correct states along the genome, we use the Viterbi algorithm (Viterbi 1967) on the final read count density (Figure 14B). The Viterbi algorithm can decode most of the states from the first run based on the estimated parameters. We noticed that for our data, predictions obtained by the first run contained a substantial amount of noise and predicted regions more narrow than we expected. To overcome these two shortcomings of the Viterbi algorithm, we introduced the iterative Viterbi algorithm, which results in predictions corresponding to longer regions containing less noise.

In the corresponding article (Ashoor et al. 2013), we demonstrated that on simulated data HMCan provides better prediction accuracy than three tools commonly used to detect histone modifications with ChIP-seq data: CCAT (H. Xu et al. 2010), MACS (Zhang et al. 2008) and SICER (Zang et al. 2009).

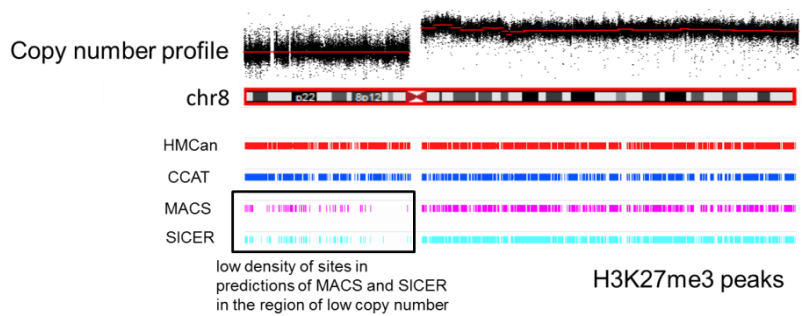

**Figure 15**. Predictions of SICER and MACS are biased toward regions of genomic gain, whereas predictions of CCAT and HMCan do no show copy number bias. Top track: copy number profile for chromosome 8 of the CL1207 human bladder transitional cell carcinoma cell line calculated by GAP (Popova et al. 2009) using SNP array technology; Bottom tracks: regions predicted to have the H3K27me3 mark by HMCan, CCAT, MACS and SICER. The black frame shows the chromosome arm 8p, which has lower density of sites predicted by MACS and SICER.

To assess the performance of HMCan on experimental data, we applied the method to the ChIP-seq data for H3K27me3 generated for the CL1207 human bladder transitional cell carcinoma cell line. First, as a result of explicit correction for copy number status, HMCan did not show any copy number bias in predictions in contrast to MACS and SICER (Figure 15). Second, we observed that gene expression values of genes with promoter H3K27me3 predicted by HMCan were significantly lower than expression values for genes with promoter H3K27me3 predicted by any of the three other methods. Since the H3K27me3 is a repression mark, this observation confirmed the better prediction accuracy of HMCan. The qPCR validation of several genomic regions in the CL1207 cell line also suggested the higher performance of HMCan compared to the other methods.

By this work, we demonstrated that the use of peak calling tools designed for normal diploid cells may lead to systematic bias when applied to cancer cell data. This bias consists of under-detection of regions with histone modifications within genomic losses and false positive calls of histone modifications within genomic gains. We observed this bias in ENCODE cancer cell data (T. E. P. Consortium 2012) for different histone modifications. Since inaccurate mapping of histone modifications in cancer cells may lead to incorrect biological conclusions, the ENCODE cancer cell line data should be reanalyzed to eliminate the copy number bias we have detected.

For this project, PhD student Haitham Ashoor designed the algorithm of HMCan, implemented the algorithm in a C++ code, tested HMCan and other computational methods on the simulated data, data for H3K27me3 generated in Institut Curie and data publicly available in ENCODE. The major part of this work was performed during his internship in summer 2012. Haitham Ashoor and I wrote a paper together during 2013. This work was published in August 2013 with me as the last author (Ashoor et al. 2013) and accepted for a talk at RECOMB'2013 and ISMB'2014.

# 3 REARRANGEMENTS OF THE GENOMIC SEQUENCE AND THEIR ROLE IN DEREGULATION IN CANCER

In this chapter, I focus on the changes at the semantics level in the cancer genome, i.e. on the genomic mutations and structural variants. Such alterations often change the meaning of some "words" and "phrases" (regulatory regions and genes) in the genomic text. This results in changes in gene expression, loss of function of genes and sometimes (in the case of mutations in proto-oncogenes) in a formation of a protein with modified function (gain of function mutation).

There is no cancer genome without alterations. These alterations can be causative but more often they are passenger events. They can be small (point mutations, indels) or large (large structural alterations). Some cancer types have preferences for small mutations (melanoma) and some for large rearrangements (neuroblastoma) (Vogelstein et al. 2013).

During my work I mostly focused on the detection of large genomic rearrangements. This is the subject of this chapter.

Large genomic changes include whole chromosomal gains and losses, translocations, large deletions, tandem duplications, insertions, inversions and amplifications. Some of these events can be seen at the 24-color spectral karyotyping (SKY) (Trask 2002): Figure 16. These alterations may influence cell functioning and result in cancer development and progression.

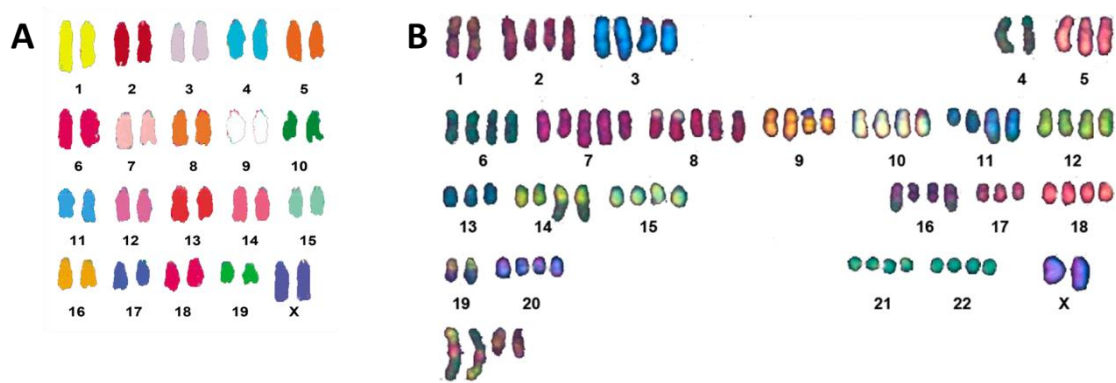

**Figure 16**. Karyotype of normal vs cancer cells. **A**. SKY karyotype of an apparently normal diploid metaphase from *p53*$^{-/-}$ mouse cell (Min et al. 2007). **B**. SKY karyotype of human neuroblastoma cell line CLB-RE (Schleiermacher et al. 2003). This cell line shows whole genome amplification, and a range of interchromosomal rearrangements, among them balanced translocation t(8;10) and unbalanced translocations t(2;3), t(9;20), t(6;11), t(4;14), t(6;16), t(11;19), etc.

Large genomic alterations may result in the formation of chimeric genes. Often, chimeric genes are not expressed at the detectable level (Boeva et al. 2013); and even expressed, the corresponding proteins seldom have an oncogenic function. However, there are classic examples when just one translocation forming a chimeric gene is enough for cancer development. One such translocation discovered in 1986 by Ben-Neriah *et al.* (1986) between genes BCR and ABL, located on chromosomes 22 and 9, results in a chimeric gene causing the development of chronic myelogenous leukemia (CML). Another example is given by Ewing sarcoma. Cancer cells of 85–90% of Ewing patients contain a specific fusion between chromosomes 11 and 22. This fusion forms a chimeric oncogene EWS/FLI1 causing this bone tumor (Delattre et al. 1992).

Large and small deletions may also play a role in tumor development and progression. Small deletions may disrupt tumor suppression genes, while large deletions may delete complete gene loci (Rajaram et al. 2013). In some cancers, large genomic deletions that encompass dozens of genes including several tumor suppressors are associated with poor prognosis. For instance, neuroblastoma tumors with 1p- or 11q deletions display a poor prognostic phenotype (Carén et al. 2010; Schleiermacher et al. 2010).

Oncogenes are often amplified in different cancers. For example, around 20% of breast cancers show *ERBB2* gene amplification and, as a consequence, overexpression of the corresponding protein: ERBB2 tyrosine kinase receptor (Sircoulomb et al. 2010). Amplification of the *MYCN* gene occurs in 20% to 25% of primary neuroblastomas (Maris and Matthay 1999).

The information about gene amplifications and specific translocations in tumors can be used to correct the clinical prognosis and select the most beneficial treatment. In some cases, cancer cells bearing fusion genes can be selectively targeted. The most famous example is targeting of CML cells expressing BCR/ABL proteins with ABL-specific tyrosine kinase inhibitor imatinib (Goldman and Melo 2003). There exist targeted therapies for cells with specific gene amplifications. For instance, *ERBB2* amplified cells may be targeted with trastuzumab or lapatinib (Sircoulomb et al. 2010). In the case of the *MYCN* amplification in neuroblastoma, although the targeted treatment does not exist, it was shown that the patients widely benefit from autologous bone marrow transplantation or peripheral-blood stem-cell transplantation (Kawa et al. 1999).

These examples show that in clinics detection of copy number aberrations (CNAs) in a tumor genome of a particular cancer patient may allow predicting tumor aggressiveness and help to choose an adequate anti-cancer treatment. In biomedical research, analysis of copy number profiles coupled with the clinical data of a cohort of patients allows detection of putative tumor suppressors and oncogenes (Cancer Genome Atlas Research Network 2011).

The next sections focus of the detection of large genomic aberrations using high-throughput sequencing data. For the detection of large genomic aberrations in the tumor, we need to make a formal distinction between CNAs and structural variants (SVs). Structural variants are objects that contain information about structural transformation of the reference genome. A list of structural variants includes deletions, insertions of known and unknown DNA fragments, tandem duplications, balanced and unbalanced translocations, etc. Some of these SVs result in CNAs (chromosomal gains and losses). For example, deletions result in losses of chromosomal material, while tandem duplications result in gains. Unbalanced translocations may result both in gains and in losses. Inversions and balanced translocations are copy-neutral events; they do not correspond to any CNAs.

Using DNA sequencing data, we can detect copy number status of targeted regions and often the corresponding number of paternal and maternal alleles. The information about the exact copy number of genomic regions in cancer allows us to say which genes or exons were lost or amplified during the tumor evolution.

DNA sequencing data can be classified as whole genome sequencing (WGS), whole exome sequencing (WES) and amplicon sequencing data (Figure 17) (Grada and Weinbrecht 2013). DNA fragments can be sequenced from one or two extremities. In the former case, the sequencing is called "single end" and the sequenced dataset contains one read per DNA fragment; in the latter case the sequencing is called "paired-end" and each DNA fragment corresponds to a pair of reads. Depending on the technology used, reads can have fixed length (25-250bp, Illumina technologies) or variable length (40-5000bp, Life Technologies or Pacific Biosciences).

For WGS, the total genomic DNA is extracted from cells and sequencer-ready DNA fragment libraries are generated from genomic DNA. Paired-end sequencing is usually applied, since it provides twice more information about each DNA fragment and allows for the better read mappability. For paired-end sequencing one usually generate reads of fixed length using the most popular at the moment Illumina sequencers.

WES, also known as targeted exome capture, consists of selective enrichment of DNA libraries with fragments overlapping the exons. There exist several target enrichment strategies including in-solution sequence capture (e.g. TruSeq Exome Enrichment Kit and Nextera Rapid Capture by Illumina) and hybrid capture (NimbleGen by Roche). The target regions in the human genome cover several dozen Mb (e.g. 37 Mb by The Nextera Rapid Capture Exome by Illumina and 62 Mb by Nextera Rapid Capture Expanded Exome by Illumina). In addition to coding exons, in the "expanded" exome one usually includes untranslated regions (UTRs) and miRNA. In WES, paired-end sequencing is more often employed than single end sequencing. Since the probes (95bp, in Illumina protocols) are located in the exons, the resulting coverage has a characteristic symmetric "bell curve" shape with higher coverage depth on exons and lower coverage depth on flanking regions (Figure 17B).

Amplicon sequencing consists of the PCR amplification of a limited number of the genomic regions of interest (amplicons) followed by high-throughput sequencing (Beadling et al. 2013). The number of amplicons varies from several hundred to several thousand. Each amplicon often coincides with an exon; exons longer than the typical length of PCR reaction products may be covered by two or more amplicons (Figure 17C). In contrast to the WGS and WES, here one usually employs single end sequencing. Currently, in amplicon sequencing

DNA fragments are sequenced with the low input/low output sequencing technology (e.g. Ion Proton by Life Technologies). The sequencers of this type are the most appropriate for sequencing small amount of reads (~1-2 millions) at low cost and high speed. Reads sequenced with these technologies have various lengths. Since the total length of the targeted regions is relatively small (usually less than 2 Mb), the mapping of reads is fast and unambiguous. The user can use any mapping algorithm that allows mapping of reads with various lengths. Due to the PCR amplification procedure used to amplify the regions of interest, one observes sharp boundaries in coverage depth with high coverage regions corresponding to the amplicons (Figure 17C).

All three types of DNA sequencing techniques are suitable for the detection of CNAs. However, one should keep in mind that while WGS provides copy number information about any genomic region present in the reference genome, using WES and amplicon sequencing one can only detect the copy numbers of the targeted part of the genome.

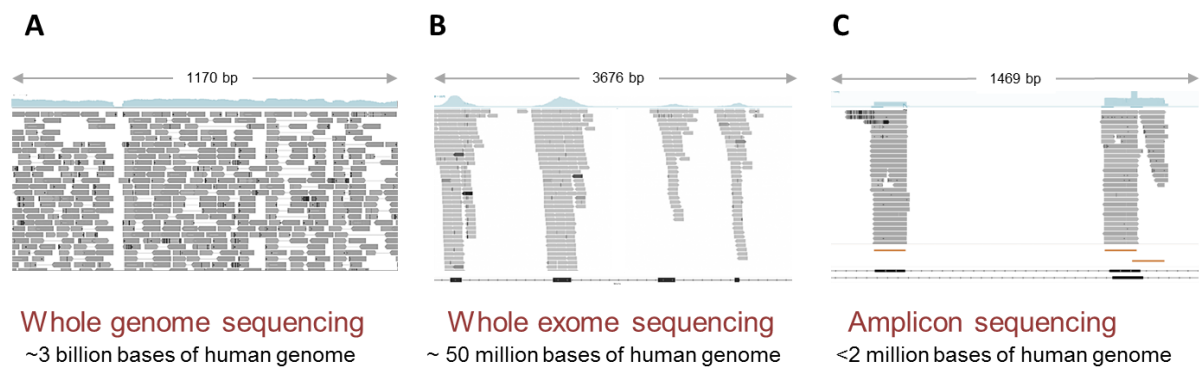

**Figure 17**. Mapping signature of three different types of DNA sequencing data. **A**. Whole genome sequencing (WGS), paired-end library; **B**. Whole exome sequencing (WES); **C**. Amplicon sequencing. The bottom track in (A) and (B) shows the exon-intron structure of the corresponding gene. The three orange bars above the exon-intron track in (C) correspond to the amplified regions (amplicons). One can see a much higher number of reads starting at the same position (duplicate reads) in the amplicon data (C) compared to the WGS and WES data (A, B). Depth of coverage is shown in blue.

The genome-wide detection of SVs is only possible using WGS data. With WEG data, an SV can be detected if both corresponding breakpoints fall within the targeted part. One targets at most 3% of the genome. This gives the probability of detection of an SV less than 0.1% given the uniform distribution of the breakpoints in the genome. With amplicon sequencing, one can only detect SVs, primers for which were included into the set of primers for PCR amplification. In the following sections, I discuss methods for CNA detection using all three types of DNA sequencing data and methods for SV detection using only WGS data.

### 3.1. Detection of copy number aberrations in cancer genomes

The computational methods for the detection of CNA from sequencing data are based on the detection and annotation of changes in the read count (RC) profiles. RC corresponds to the number of reads, read pairs or read starting positions per genomic region. The higher the number of copies of a given region physically present in the cell nucleus, the higher the number of sequenced reads mapping to this genomic region. Intuitively, high RC should correspond to copy number gains while low RC should correspond to copy number losses.

An appropriate definition of RC is very important and depends on the sequencing technique. In WGS data, RC often refers to the number of reads starting in a given window. The size of the window, and thus, the resolution of the breakpoint detection, depend on the total number of reads and the size of the reference genome (Boeva et al. 2011). In WES, RC corresponds to the number of read mappings overlapping each exon (Amarasinghe, Li, and Halgamuge 2013); however, the window approach can be also applied. In amplicon sequencing, RC is calculated per amplicon. If an exon is targeted using two or more closely located or overlapping amplicons, each read is assigned to only one amplicon region, the one with which the read alignment has the maximum overlap.

The further analysis of RC profiles to obtain absolute or relative copy number profiles includes normalization, segmentation and annotation of the segmented genomic regions. The concrete realization of each step depends on the experimental design and biases specific to each sequencing protocol.

### 3.1.1. Detection of copy number aberrations with WGS and WES data

Many computational methods have been published that allow automatic calculation and analysis of copy number profiles from WGS and WES data (Chiang et al. 2009; Xie and Tammi 2009; Yoon et al. 2009; Ivakhno et al. 2010; Abyzov et al. 2011; Lonigro et al. 2011; Sathirapongsasuti et al. 2011; Xi et al. 2011; Carter et al. 2012; Fromer et al. 2012; Klambauer et al. 2012; Koboldt et al. 2012; Krishnan et al. 2012; Amarasinghe, Li, and Halgamuge 2013; Magi et al. 2013; Mayrhofer, DiLorenzo, and Isaksson 2013; Oesper, Mahmoody, and Raphael 2013; Yau 2013). The first methods (Xie and Tammi 2009; Chiang et al. 2009) applied normalization of reads counts in the tumor samples with the matched normal control dataset, followed by segmentation and statistical tests to detect CNAs. The latest tools allow for extensive RC normalization and may work without matched normal control (Klambauer et al. 2012). In addition, the most recent tools use B allele frequencies calculated from read coverage of SNP positions to assess the level of contamination of tumor samples by normal cells and the presence of subclones, and thus detect the absolute number of copies of each region (Mayrhofer, DiLorenzo, and Isaksson 2013; Oesper, Mahmoody, and Raphael 2013). When the matched control sample is present, the current tools can discriminate between germline and somatic CNAs present in cancer genome (Koboldt et al. 2012; Krishnan et al. 2012; Yau 2013).

We were pioneers in developing CNA detection methods for WGS and WES data. Our method, FREEC, was the first method to analyze tumor samples without a matched normal control, to take into account tumor ploidy and correct for tumor contamination by normal cells (Boeva et al. 2011). One year later, we extended FREEC to take into account B allele frequency (BAF) information in order to predict genotypes genomic regions and be able to detect copy-neutral loss of heterozygosity (LOH) regions (Boeva, Popova, et al. 2012).

Since then, FREEC performance was assessed in different independent comparisons of CNA calling methods (J. Duan et al. 2013; Janevski et al. 2012; Krishnan et al. 2012). For instance, Duan et al. say that when high true positive rate is preferable, FREEC is one of the best choices among six publicly available CNA detection methods. Overall, FREEC became a standard CNA calling tool for cancer genome analysis. In addition to the tumor data, FREEC is now widely applied to non-human data for the detection of CNA in genomes of model and non-model organisms (Doan et al. 2012). FREEC was integrated in the tumor data analysis pipeline in Institut Curie and is accessible for external use via command line (http://bioinfo-out.curie.fr/projects/freec/).

In FREEC we developed methodology to overcome the following challenges: read count dependency from the percentage of G/C in a region (GC-content bias), read count dependency on the uniqueness of the corresponding region (read mappability bias) and changes in the expected read count and BAF related to contamination by normal cells.

GC content bias arises from the PCR amplification step done prior to sequencing (Dohm et al. 2008). The PCR amplification distorts the RC in a very considerable measure (Figure 18). We showed that when the matched normal control is not available, the GC-content can be used to normalize the RCs. In FREEC, we use mappability profiles generated by GEM (Faust and Hall 2012) and excluded from the analysis low mappability regions. The user can also choose an option to correct for low mappability.

In order to detect CNAs, the normalized RC profiles should be segmented. Segmentation allows for a much more accurate detection of regions of gain and loss than point-by-point assessment of copy number status based on a predefined threshold. Different segmentation techniques can be applied. The most commonly used are HMMs (Amarasinghe, Li, and Halgamuge 2013) and circular binary segmentation (Miller et al. 2011). In FREEC, we implemented a LASSO-based algorithm suggested by (Harchaoui and Lévy-leduc 2008).

In cancer, copy-neutral LOH regions often contain mutated copies of tumor suppressor genes. Copy-neutral LOH, also called acquired uniparental disomy (UPD) or gene conversion, arises when a maternal (paternal) chromosome or a chromosomal region is duplicated when the paternal (maternal) counterpart is lost. Both copies of tumor suppressor genes in such regions may be disabled by identical point mutation or indel.

The extension of FREEC – Control-FREEC – estimates BAFs from nucleotide coverage of polymorphic positions and detects genotypes of genomic regions including copy-neutral LOH (Boeva, Popova, et al. 2012). To estimate BAFs, we calculate the proportion of reads mapping to the alternative allele B instead of the reference allele A for the known SNP positions (Figure 19A). The resulting BAF profiles are then segmented and annotated using Gaussian mixture model (GMM) fit with fixed means (Figure 19B). We calculate the likelihood of fit for several sets of fixed mean values; these values depend on copy number status $N$ of a given region and normal contamination $c$. The mean values for a tested number of alternative alleles $k$ are calculated as follows:

$$\text{mean for } (A)_{N-k}(B)_k = \frac{k(1-c)+c}{N(1-c)+2c}$$

$$\text{mean for } (A)_k(B)_{N-k} = 1 - \frac{k(1-c)+c}{N(1-c)+2c}$$

$$\text{means for } (A)_N = \begin{cases} 0.11, & \text{when testing for } k > 0 \\ 0.11 \text{ and } \max\left(0.11, \frac{c}{N(1-c)+2c}\right), & \text{when testing for } k = 0 \end{cases}$$

$$\text{means for } (B)_N = \begin{cases} 0.89, & \text{when testing for } k > 0 \\ 0.89 \text{ and } \min\left(0.89, 1 - \frac{c}{N(1-c)+2c}\right), & \text{when testing for } k = 0 \end{cases}$$

In the corresponding paper (Boeva, Popova, et al. 2012), we showed that our results on ~30x-coverage whole genome sequencing dataset agreed with the SNP-array analysis output. We obtained 95.4% consistency between the results of Control-FREEC and GAP applied to SNP array data generated for the same tumor sample (Popova et al. 2009).

In addition to WGS, FREEC can be applied to WES data. Because of additional capture bias, the normalized copy number signal in WES is often more of lower quality than from the signal from WGS. Thus, we provided an option to correct copy number predictions using BAF GMM fit.

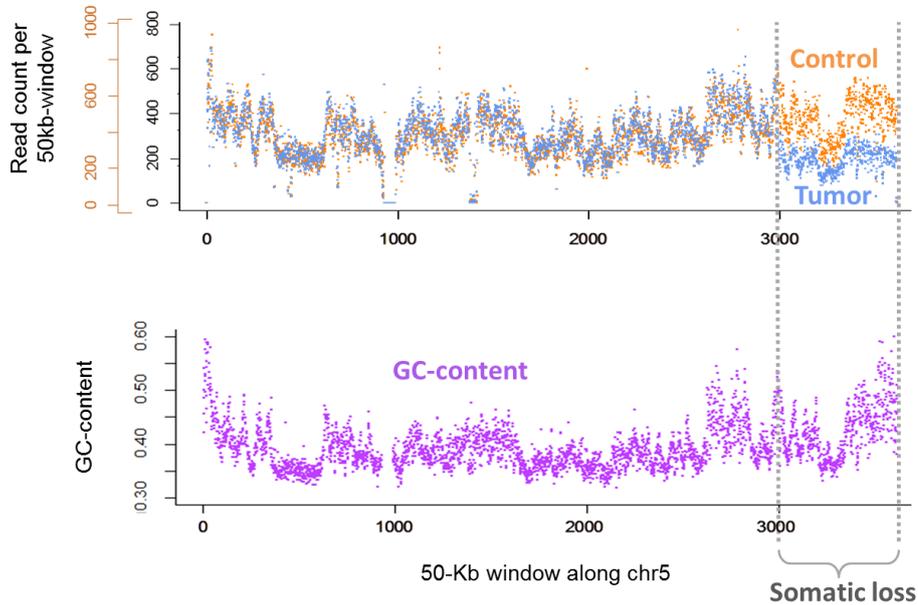

**Figure 18**. Dependency between the read count (RC) and GC-content. RC corresponds to the number of reads starting in a 50-Kb window: neuroblastoma cell line CLB-GA (blue) and diploid control (orange). GC-content corresponds to the proportion of nucleotides C and G in a given window (purple).

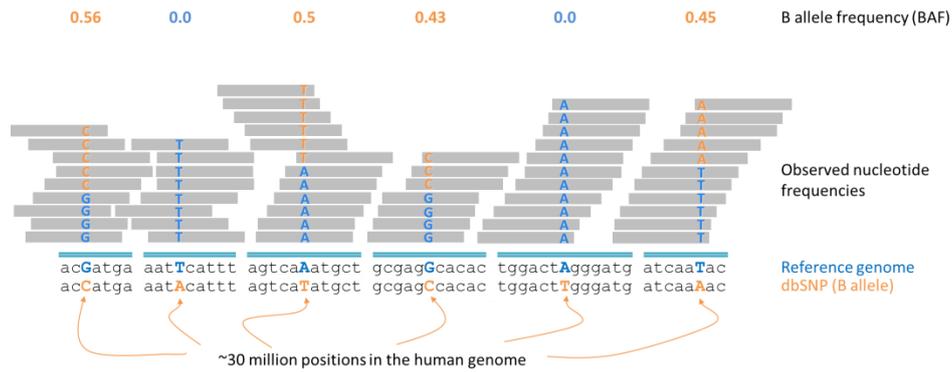

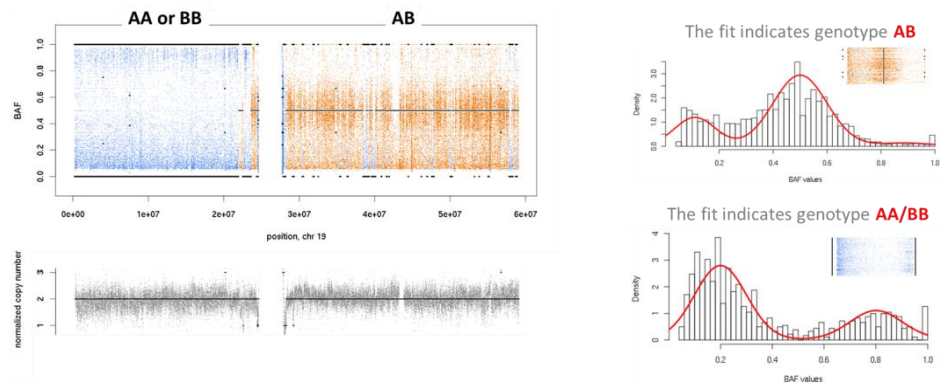

**Figure 19**. Detection of genotype and loss of heterozygosity (LOH) regions. **A**. Calculation of B allele frequency (BAF) profile using the read coverage of the reference and alternative allele. Reads are shown by grey rectangles; nucleotide corresponding to the alternative allele on SNP positions ar shown in orange. The observed BAF value at a given SNP position is the percentage of reads mapping to the alternative allele among all reads covering the given SNP. **B.** BAF (blue and orange) profile and copy number profile (grey) for chromosome 19, 30x-coverage WGS neuroblastoma dataset. LOH and heterozygous regions are shown in blue and orange respectively. Predicted BAF and copy number profiles are shown in black.

### 3.1.2. Detection of copy number aberrations with amplicon sequencing data

While relatively expensive WES consists of in-depth sequencing of nearly all the coding exons, the amplicon sequencing technique aims at sequencing a limited number of genes (from several dozen to several thousand exons) at an extremely low cost. The genes included in a panel of amplicon sequencing (actionable genes) are genes that are often altered in different cancer types, and for whose alterations targeted therapies have been established or are in clinical development. For instance, the TargetRich™ CRX kit from Kailos Genetics assays such cancer-related genes as BRAF, EGFR, FLT3, JAK2, KIT, KRAS, PIK3CA, PTEN, TP53 and VEGFA; the AmpliSeq™ Cancer Panel from Life Technologies targets 190 regions of interest in 46 well-characterized oncogenes and tumor suppressors. Some actionable genes often undergo point mutation or exon deletions (e.g. ALK, BRAF) while others undergo amplification in copy number (e.g. MYCN, ERBB2) (Garraway and Lander 2013; Small et al. 1987). Cancer cells with amplifications of certain genes can be specifically treated with targeted therapies. For example, a personalized treatment is available for patients with amplification of the *ERBB2* gene (approved in clinical practice for breast and gastric cancer patients): the extracellular domain of ErbB2 can be targeted by trastuzumab or pertuzumab, while the intracellular domain (tyrosine kinase) can be inhibited by lapatinib (Tai, Mahato, and Cheng 2010).

There are two major differences in RC profiles in amplicon sequencing and WES data. First, there is a much higher noise in amplicon sequencing compared to exome sequencing due to a more complex library preparation technique. Second, amplicon sequencing targets fewer regions and thus provides less information than exome sequencing datasets (<10,000 exons vs >200,000 exons); consequently, data normalization can be less effective on amplicon sequencing data.

In collaboration with biotechnological company OncoDNA, we developed ONCOCNV, a method to identify CNAs of the actionable genes targeted by amplicon sequencing. The RC processing method implemented in ONCOCNV includes normalization for library size, GC-content of each amplicon region and amplicon length. We call these kinds of bias 'library specific'. The exact shape of the functional dependency of the RC on the GC-content and amplicon length may be different in each library. In addition to the library specific bias, there is a technology-specific bias. This bias, even after normalization for the library-specific bias, is present to a different extent in the RCs. Therefore, for each technological platform, we construct a baseline that reflects the technological bias in the diploid control samples and use it for the normalization of the test samples. To construct the baseline, we apply principal component analysis (PCA) to the normalized RCs from the diploid samples. The only requirement for the selection of diploid control samples is that they should be processed using the same target selection kit as the tumor samples. For the final normalization of the sample RCs, we use a linear regression over the baseline.

In the corresponding paper, we demonstrated that ONCOCNV significantly outperforms methods such as ADTEx (Amarasinghe, Li, and Halgamuge 2013) and NextGENe (commercial software) developed for the analysis of WES data. We did not position ONCOCNV as a tool for WES analysis, although technically it can be applied to exome data. Unlike conventional exome sequencing data analysis tools, ONCOCNV does not take into account B allele frequencies. Due to this limitation, ONCOCNV is not able to evaluate the level of the contamination by normal cells or improve the accuracy of CNA calling by simultaneously processing copy number and B allele frequency profiles.

## 3.2. Detection of structural variants in cancer genomes

The identification of genomic structural variants (SVs) is often a key step in understanding cancer development (Futreal et al. 2004). As it was mentioned in the beginning of this chapter, specific gene fusions in certain cell types can have oncogenic potential (Delattre et al. 1992; Ben-Neriah et al. 1986). SVs occurring in tumor suppressor genes can disrupt their function and thus promote oncogenesis (Min et al. 2007; Hanahan and Weinberg 2011).

With the arrival of high-throughput sequencing technologies, the use of short-insert paired-end reads enabled genome-wide detection of SVs (Korbel et al. 2007). With *a priori* information for paired-ends such as order, orientation and insert size of pairs as constraints during read alignment to the reference genome, anomalously mapped pairs indicate potential genomic variations from the reference (Figure 20). Most of the large genomic rearrangements (insertions, deletions, inversions, tandem and mirror duplications, translocations) can be detected through the analysis of the abnormal paired read mappings. For instance, a translocation is characterized by abnormal read pairs with first and second read in each pair mapping on different chromosomes; a deletion can be detected as a cluster of read pairs with abnormally high insert size (Figure 20B).

SV detection approaches based on clustering and annotation of abnormal paired read mappings were implemented in several SV calling methods (Korbel et al. 2007; Chen et al. 2009; Korbel et al. 2009; S. Sindi et al. 2009; Wong et al. 2010), including SVDetect (Zeitouni et al. 2010) developed by our team.

For the detection of genomic rearrangements, one usually applies one of the two strategies: paired end or mate-pair sequencing. Paired end sequencing is basically sequencing of DNA molecules from both ends after DNA fragmentation. This results in inward orientation of read mappings in a read pair. Mate-pair library preparation includes DNA fragmentation, circularization of DNA fragments with simultaneous labeling of the ligation site with deoxynucleotide triphosphates, extraction of labeled circularized molecules, their fragmentation and purification of the labeled fragments (corresponding to the ends of the original DNA ligated together). Purified fragments then are sequenced using paired-end sequencing protocol. In contrast to paired-ends, mate-pair sequencing provides much larger insert sizes. Mate-pair sequencing results in outward orientation of read mappings in a read pair in Illumina protocols and in co-directed orientation (Reverse-Reverse or Forward-Forward) in SOLiD protocols. The approaches listed above can be applied to either to high coverage paired end libraries or to mate-pair libraries (usually low coverage).

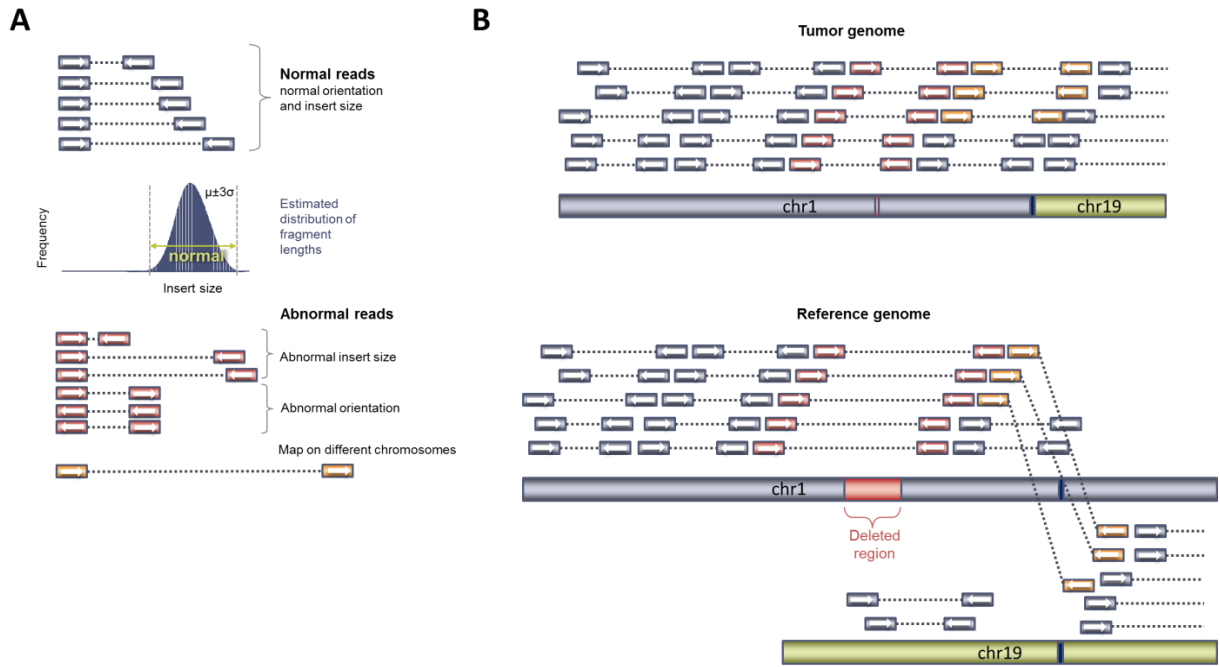

**Figure 20**. Detection of structural variants based on paired-end reads. **A**. Separation of read pairs into normal and abnormal pairs according to the mapping signature (orientation, insert size). The normal reads are reads mapping on the same chromosome with orientation specified by sequencing protocol (Illumina, SOLiD paired ends = Forward-Reverse, Illumina mate-pairs = Reverse-Forward, SOLiD mate-pairs = Reverse-Reverse/Forward-Forward) and insert size falling within the interval of expected fragment lengths. **B**. Example of abnormal reads clustering into read signature signifying deletion (red) and translocation (orange).

Other methods use an alternative approach for SV detection based on partial read alignment (Jianmin Wang et al. 2011; Schröder et al. 2014). Although they may be efficient for data with high read coverage, they may fail identifying SVs with the breakpoints located in the repetitive elements of the genome. Ideally, this type of approaches should be combined with paired-end signatures; this idea was implemented in SVMerge (Wong et al. 2010), PRISM (Jiang, Wang, and Brudno 2012) and DELLY (Rausch et al. 2012).

Combining information of paired end signatures with information about changes in read coverage depth is another promising direction. Indeed, many unbalanced SVs, such as deletions, tandem duplications, co-amplifications, unbalanced translocations, result in copy number alterations and thus in changes in read depth. Probabilistic models integrating both the read depth signal and paired-end signatures provide higher specificity at equal or greater sensitivity than tools that simply use paired-end signatures (Oesper et al. 2012; S. S. Sindi et al. 2012; Escaramís et al. 2013).

Our method SVDetect (Zeitouni et al. 2010) was one of the first methods for SV detection among those using only paired-end signatures. This method is still widely used by the research community as it has a number of features lacking in the more recent tools.

SVDetect contains rules for the detection of signatures of a large spectrum of genomic rearrangements. It includes deletions, insertions, tandem duplications, amplifications, co-amplifications, inversions, fragment reinsertion, inverted duplications, balanced and unbalanced translocations. SVDetect identifies and annotates clusters of abnormal reads inconsistent with any SV signature. Such clusters are often due to incorrect mapping of reads coming from repetitive genomic elements. Also, SVDetect is able to simultaneously process several samples and annotate common and sample-specific SVs. In cancer studies, it allows filtering out germline events. Importantly, SVDetect is able to process data coming from various sequencing protocol and technologies; it accepts both mate-pair and paired end reads of Illumina and SOLiD sequencing machines.

### 3.2.1. From computational strategies to the discovery of chromothripsis in neuroblastoma

Characterization of genomic rearrangements in neuroblastoma performed in collaboration with the laboratory of Genetics and Biology of Cancers in Institut Curie provides an example of application of the methods for SV and CNA discovery presented in the previous sections.

Neuroblastoma, the most frequent extracranial paediatric solid tumor, accounts for 8-10% of deaths from cancer in childhood and is characterized by a great clinical and genetic heterogeneity. Several types of somatically acquired chromosomal imbalances have been described in this cancer. These include whole chromosome gains or losses, associated with ploidy abnormalities, and structural chromosome alterations, occurring as genomic amplifications or unbalanced translocations (Brodeur 2003).

Whereas at the constitutional level, a few translocations have been characterized at the gene and base pair level in neuroblastoma patients, only one unbalanced somatic translocation has been explored at the base pair level in sporadic neuroblastoma (Schleiermacher et al. 2005).

The goal of our study was to characterize at the base-pair level two neuroblastoma cell lines and two neuroblastoma primary tumors using mate-pair sequencing.

The copy number status of each genomic region was assessed using FREEC (Boeva et al. 2011), while putative SVs were predicted using SVDetect (Zeitouni et al. 2010).

SVDetect yielded thousands of candidate SVs, most of them were false positives. Using the annotations provided by SVDetect, it was possible to filter out most of these false positive calls. We applied the following filters on the candidate SVs (links): (1) minimal number of pairs/link, (2) minimal proportion of consistent pairs; we removed links (1) connecting telomeric/centromeric regions, (2) falling in satellite sequences of the same type, (3) falling in segmental duplication regions, and (4) links detected both in the tumor and normal sample.

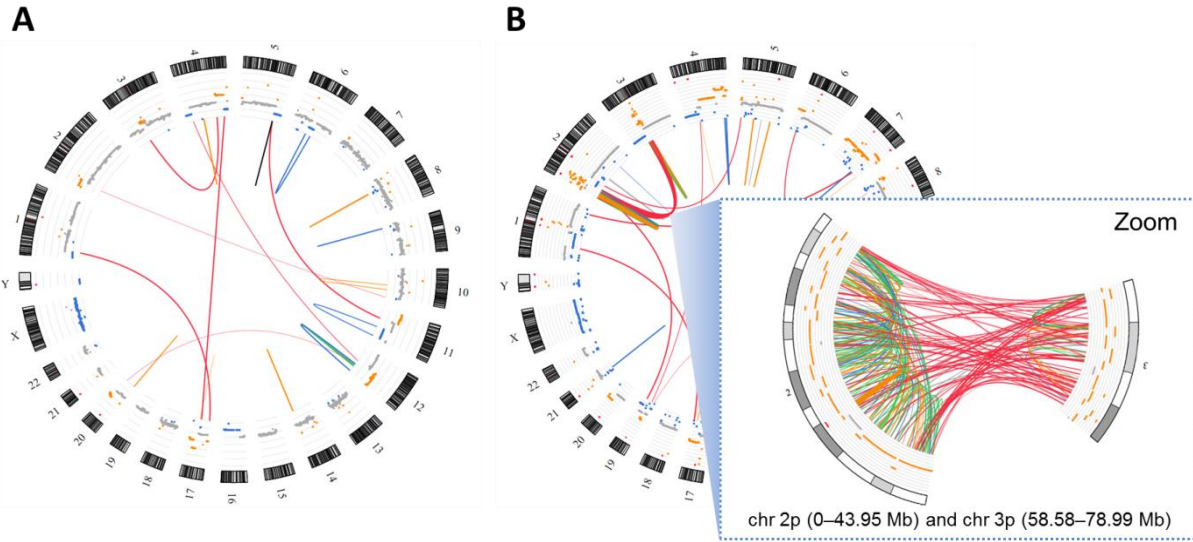

**Figure 21**. Genome-wide profile of predicted SVs in CLB-Ga (**A**) and CLB-Re (**B**) cell lines. All inter-chromosomal rearrangements and intrachromosomal SVs encompassing regions longer than 50 kb identified in mate-pair sequencing are visualized using Circos (Krzywinski et al. 2009). Chromosomes appear as ideograms. The outer ring shows a representation of copy number as determined by sequencing data (grey: normal copy number; blue: deletion, orange: gain). The inner circle shows the two endpoints of each rearrangement identified (black: inverted duplication, blue: deletion, green: inversion, orange: large duplication, red: unbalanced inter-chromosomal rearrangements). Most of the identified unbalanced rearrangements correspond to unbalanced translocations previously identified by 24-color karyotyping. Line's thickness is related to the numbers of pairs identified in each link. The dense clustering of SVs was unexpected considering previous characterization of the rearrangements present in this cell line by array-CGH and 24-color karyotyping.

In the cell lines, most of the inter-chromosomal rearrangements detected by SVDetect matched the unbalanced translocations previously detected by spectral karyotyping and array-CGH. However, in one of the cell line (CLB-RE), our approach unexpectedly detected hundreds of intra- and inter-chromosomal rearrangements (Figure 21B). These SVs were localized within chromosomes 2 and 3; the pattern of rearrangements was highly reminiscent of chromothripsis, recently described in several cancer types (Korbel and Campbell 2013). Chromothripsis is characterized by massive genomic rearrangement acquired in a single catastrophic event during cancer development (Stephens et al. 2011). Although there are several theories about the origins of chromothripsis, the exact molecular mechanisms of chromothripsis remain unknown (Holland and Cleveland 2012; Sorzano et al. 2013).

The observation of chromothripsis in one of the cell lines prompted us to investigate two primary NB tumors presenting with shattering of one or two specific chromosomes, previously detected by array CGH. Mate-pair analysis confirmed the geographic localization of the rearrangements and documented their diversity. Using complementary experiments by PCR and Sanger sequencing, we characterized 51 rearrangements in the four samples at the base pair resolution that revealed 59 junctions. In a subset of cases, complex rearrangements were observed with templated insertion of fragments of nearby sequences. A detailed inspection of the breakpoints revealed frequent microhomologies at the junction points. Our study supported the hypothesis that not only Non-Homologous End Joining (NHEJ)-mediated repair but also replicative processes such as Break-Induced Replication (BIR), Fork Stalling and Template Switching (FoSTeS) and/or Microhomology-Mediated Break-Induced Replication (MMBIR) may account for genomic rearrangements in neuroblastoma in both chromothripsis and non-chromothripsis cases.

Also, we identified several structural variants targeting genes located within common fragile sites. The FHIT gene was targeted by multiple rearrangements in one sample; FHIT overlies the fragile site FRA3B. The WWOX gene overlapping the FRA16D site was implicated in another sample with two rearrangements.

RNA-seq experiments confirmed expression of several predicted chimeric genes and genes with disrupted exon structure. Genes with abnormal transcripts included ALK, NBAS, PTPRD and ODZ4. These four genes have been shown to be frequently mutated or targeted by rearrangements in other neuroblastoma samples (Molenaar et al. 2012).

# 4 CONCLUSIONS AND PERSPECTIVES

Deciphering regulation in eukaryotic cells becomes increasingly easier with the emergence of novel techniques generating large amount of high-throughput data, in particular, high-throughput sequencing (HTS) data. These data provide genome wide information about DNA mutations, chromosomal rearrangements, coding gene and small RNA expression, DNA-protein binding, epigenetic changes, 3D-structure of the genome and many other molecular phenomena.

Going from sequence to function using these data is still a complicated task. There are two levels of challenges in high-throughput data analysis: data processing and data interpretation.

The open questions in data processing are very much related to the data normalization and the extraction of the biologically meaningful signal. Data interpretation, being much wider than data processing, employs knowledge from many scientific areas, including molecular biology, statistics, theory of probability, and computational learning theory. The proper data analysis often combines elements of sequence analysis, supervised and unsupervised classification, graphical models, etc. Data interpretation projects are driven by a biological question. Ideally, they should imply a back and forth collaborative process when the analysis of data inspires new experiments, and new experiments provide more data for the bioinformatics analysis. In addition to the integration of scientific methods during data interpretation, researchers often have to integrate different data types. Although data integration has recently become a hot subject in life sciences and biomedical research, there is no simple "recipe" that could be applied everywhere, and each particular biological question needs its proper data integration strategy.

My vision is that the development of novel computational techniques for data processing and data interpretation should go hand-in-hand with deep data analysis. Useful computational methods are likely to be developed when one has a clear goal – a clear biological question to answer. My ambition is to develop novel computational techniques for analysis of cancer HTS data and in parallel contribute to our understanding of the mechanisms of transcriptional regulation and exploit epigenetic changes in cancer cells.

*Data processing challenges*

*Data normalization.* At the level of data processing, one of the main challenges in HTS data analysis is normalization of the signal (usually read count). Depending on the experiment, the read count characterizes the level of RNA transcripts, the number of DNA copies, the binding affinity of DNA associated proteins, the amount of physical contacts between different genomic regions, etc. The signal contained in the read count may be distorted by

(i) technical biases due to data generation processes: e.g. GC-content bias in most of the HTS applications involving PCR amplification, amplicon size bias in amplicon sequencing, increased read count in "open chromatin" regions compared with "repressive chromatin";
(ii) read mappability: e.g. lower/higher read count in repetitive regions when one consider unique/multiple matches.

*Characterization of the signal coming from the repetitive elements.* It is extremely common to focus the study on the unique part of the genome by considering only uniquely mapped reads. Yet, the information about the signal coming from the repetitive elements is valuable. For instance, the number of genomic repetitions in tandem-repeat polymorphisms can be associated with phenotypic traits including various brain disorders, Huntington's disease, various ataxias (Hannan 2010), and cancer (T.-M. Kim, Laird, and Park 2013). Also, with cancer were associated structural variants in transposable elements (Lee et al. 2012). Furthermore, it is important to study the epigenetic landscape of repetitive elements. For example, it may allow understanding of the mechanisms of regulation of centromeric and telomeric chromatin states (Lacoste et al. 2014) and changes of chromatin state of repetitive elements due to the external stimulus (Maze et al. 2011). So far, there exist very few methods to characterize the signal coming from the repetitive elements of the genome (Alkan et al. 2009; Chung et al. 2011; Jie Wang et al. 2012, 119). We need to further develop novel computational methods to solve this important question.

*Single cell data.* If until now, we studied transcription and genomic variation at the level of cell population, then now, with single cell sequencing technologies, we start to see differences between cells *within* a cell population. The first such studies included single cell transcriptomics experiments (Islam et al. 2014; Jaitin et al. 2014) and DNA sequencing following whole genome amplification (Navin et al. 2011; X. Xu et al. 2012; Hou et al. 2012). Technologies for single-cell amplification and sequencing are maturing (Korfhage et al. 2013). It is extremely likely that soon we will be able to sequence DNA from single cells without amplification step bias and investigate protein-DNA binding at the single cell level. Analysis of single cell data is already shifting our paradigm about the molecular processes happening in the cell. For instance, single-cell RNA sequencing experiments have already revealed dynamic, random monoallelic gene expression in mammalian cells (Deng et al. 2014). With these novel technologies, we start asking new biological questions. We can study clonal evolution in cancer (Navin et al. 2011). Simultaneous sequencing of single cell genome and transcriptome will soon enable evaluation of the effects of DNA variants on transcript levels in single cells.

With the analysis of single cell data, the researchers meet a new challenge. Single cell techniques are more error-prone than widely used next- (or second-) generation sequencing techniques. Also, the amplification step routinely applied in current single cell data distorts the signal in the technology-specific manner (Kharchenko, Silberstein, and Scadden 2014). Thus, there is an increased need for methods to analyze data coming from single cell sequencing. We shall develop new types of statistical approaches to deal with such data.

*Novel HTS applications.* Bioinformatics community made already a great effort to develop methods for the analysis of "classic" HTS applications: WGS, WES, ChIP-seq, RNA-seq, and small RNA sequencing. However, every year researchers design novel applications of HTS that need development of new data processing methods. In 2008, MeDIP-seq was proposed as a method to estimate absolute DNA methylation levels along the genome (Down et al. 2008). Also, the year 2008 brought us GRO-seq, a method that allows mapping the position, amount, and orientation of transcriptionally engaged RNA polymerases genome-wide (Core, Waterfall, and Lis 2008). The same year, HITS-CLIP was developed to identify functional protein-RNA interactions *in vivo* (Licatalosi et al. 2008). In 2009, genome-wide analysis of translation with nucleotide resolution became possible using ribosome profiling (ribo-seq) (Ingolia et al. 2009). Also in 2009, the ChIA-PET strategy was published for the *de novo* detection of global chromatin interactions mediated by a DNA binding protein of interest (Fullwood et al. 2009). In 2010, HITS-CLIP was followed by PAR-CLIP, designed to detect transcriptome-wide the binding sites of RNA-binding proteins and miRNA-binding proteins at higher resolution (Hafner et al. 2010). Also in 2010, RIP-seq was developed to capture and sequence RNA molecules bound by the Polycomb repressive complex 2 (PRC2) (Zhao et al. 2010). Techniques to study 3D conformation of chromatin based on chromosome conformation capture (3C) technique developed by Job Dekker et al. (Dekker et al. 2002) and HTS started to appear in 2010 (4C) (Z. Duan et al. 2010), followed by 5C (Ferraiuolo et al. 2012) and Hi-C (Belton et al. 2012). The year 2011 brought use ChIP-exo, a method to detect genome-wide protein-DNA interactions at single-nucleotide resolution (Rhee and Pugh 2011). The years 2012 and 2013 ware marked by the transcriptome-wide characterization of human circular RNAs (Salzman et al. 2012; Memczak et al. 2013; Jeck et al. 2013). This list will continue to grow, which will require from the bioinformatics community developing novel data processing methods.

*HTS applications to new species.* Often, a method developed to analyze data for a certain type of organism (e.g. diploid human cells) may bring incorrect results when applied to other organisms (e.g. highly repetitive polyploid plant genomes). Indeed, the proportion of the uniquely mappable genome in eukaryotes is relatively high compared to the plant species. In the human genome, which contains 45% of repetitive sequences, only 27% of the genome is not uniquely mappable with 75bp reads allowing 2 mismatches (Derrien et al. 2012). While, for example, the hexaploid wheat genome contains more than 80% of repetitive sequences (Flavell et al. 1974), and more than 50% of the transcriptome is not uniquely mappable with 120bp reads (Trick et al. 2012). Novel algorithms should be developed, or corrections to the existing solutions should be applied, in order to enable proper analysis of the HTS data coming from the polyploid and highly repetitive genomes.

*HTS applications to cancer.* Unluckily for cancer research, most of the data processing methods have been developed without taking into account the specificities of cancer data. To extract the true signal from a cancer high-throughput dataset, we need to consider intratumoural heterogeneity, possible contamination by normal cells as well as different ploidy and copy number alterations. Also, novel dedicated methods are often needed for the analysis of cancer genome data in order to respond to cancer-specific questions (e.g. detection of abnormal transcripts in RNA-seq data).

Currently, three PhD students work with me on my projects involving the development of methods for analysis of HTS data produced from tumor samples:

- Detection of differential histone modifications using ChIP-seq data generated for genomes with copy number abnormalities. H. Ashoor (King Abdullah University of Science and Technology, Saudi Arabia). Co-supervision with Pr. Vladimir B. Bajic.
- Detection of structural variants in cancer genomes using Bayesian inference. D. Iakovishina (AMIB INRIA, LIX Ecole Polytechnique, France). Cotutelle with Dr. Mireille Régnier.
- Dynamic indexing of DNA sequences with application to mutation analysis in cancer cells. K. Břinda (University Paris-Est Marne-la-Vallée, France). Co-supervision with Dr. Gregory Kucherov.

*Data interpretation challenges*

Studying data in several dimensions – epigenetic changes, gene expression, copy number changes, SNPs and mutations, transcription factor binding – is beneficial for understanding the complete picture of deregulations happening in cancer. In particular, these data, assembled for a large cohort of patients and coupled with clinical data, can elucidate the cancer-driving mechanisms. Integration of different types of high-throughput data is a methodological and computational challenge.

In the following paragraphs I mention several directions involving integration of different types of HTS datasets that I plan to investigate.

*Modeling the effect of cis-regulatory elements on the modulation of gene expression in normal cells and cancer.* Although the enhancer function does not depend on the enhancer orientation, the number of TFBSs within enhancer, their orientation and distance between them affect the extent of gene expression modulation (Ridinger-Saison et al. 2012; Andersson et al. 2014). The ChIP-seq TFBS data coupled with gene expression data can be exploited to construct a predictive model for the transcriptional effect of a given combination of TF binding motifs. Resolving the model will (1) provide the knowledge about complexes of interacting TFs in the given cell type; (2) it will picture the spatial organization within the protein/DNA complex and delineate physical constrains; (3) it will clarify why some genes are activated by the TF under study only in a limited number of cell types. The model can be constructed using machine learning approaches such as support vector machines (SVM) or random forests. The input vectors can correspond to cis-regulatory regions. They can contain information about presence/strength of motif hits for the considered transcription factors, distance to gene TSSs, pair-wise distances between oriented motif hits, etc.

*The role of epigenetic remodeling in cancer.* By combining ChIP-seq histone modification data and copy number profiles, one can study aberrant epigenetic gene remodeling in cancer. One of my ongoing projects in collaboration with the Institut Curie Unit of Genetics and Biology of Cancers aims to investigate whether neuroblastoma oncogenesis may be partially explained by epigenetic silencing or activation of genes. In order to determine regions of silenced chromatin, ChIP-seq experiments with antibody specific for H3K27 trimethylation (H3K27me3), which is a histone mark associated with binding of polycomb repressor complex 2 (PRC2), will be performed on neuroblastoma cell lines. We plan to determine whether long range epigenetic silencing (LRES) can be observed in such cell lines, i.e., whether large clusters of consecutive genes can be repressed by epigenetic mechanisms in neuroblastoma. Since neuroblastoma tumor suppressor genes are potentially located in regions of common chromosomal loss, we would like to investigate whether any specific genes in these regions show frequent patterns of epigenetic silencing. Also, we plan to investigate possible correlation (or anticorrelation) between genomic losses and epigenetic silencing in neuroblastoma cell lines that are derived from aggressive primary tumors. If epigenetic silencing mechanism is confirmed in neuroblastoma, it will be crucial to study the affected pathways and its relationship with the aggressiveness of neuroblastoma. This could be addressed using neuroblastoma primary tumors of various stages and genomic profiles.

*The role of subclonal mutations in cancer progression.* Characterization of the evolution and impact of subclonal mutations in cancer can be assessed using the combination of high depth WES, amplicon sequencing and RNA-seq. In my collaborative neuroblastoma project, the analysis of high coverage WES data (100x) will allow us to obtain an overview of all genetic changes present in the tumor samples at the time of diagnosis and at relapse. We could get insights into the intratumoral heterogeneity of neuroblastoma at both time points, identify subclones with driver mutations that expanded over time and define "candidate genes" containing such mutations. To study the overall distribution of the mutation and CNAs in candidate genes in neuroblastoma patients and get insights into the association of these mutations with survival, we will perform high depth of coverage targeted sequencing (10000x) of the candidate genes in a cohort of 300 neuroblastoma tumors. The clinical data including the survival information and copy number profiles (array CGH) are available for this cohort. In addition, we will have transcriptome (RNA sequencing) data available for a subset of patients. The analysis of these data will allow us to see at what degree the subclonal mutations in candidate genes are expressed. We will also search for associations in overall gene expression with the presence of particular mutations in candidate genes. At the final stage, we are going to model the neuroblastoma progression based on the information acquired at the previous analysis steps. We are going to apply pathway analysis and network modeling to get insight into the functional association of the mutated genes (Hofree et al. 2013). The linking of the patterns of clonal evolution in neuroblastoma with activated or repressed signaling pathways should provide insights into neuroblastoma stepwise transformation. The fellowship for a PhD student to work under my supervision with co-supervision by Dr. Olivier Delattre was allocated by AMX Ecole Polytechnique doctoral school in June 2014.

*Changes in the genome 3D structure in cancer and its relationship to aberrant gene expression.* Such fundamental questions as changes in 3D conformation in cancer genomes due to structural variants can be assessed using the integrative analysis of LADs, WGS data, ChIP-seq for histone marks and sequence analysis of transcription factor binding motifs. Large structural variants, such as translocations, large genomic deletions or inversions, disrupt the canonical 3D structure of the genome. The genes located in regions associated to the nuclear lamina (LADs) in the cells of cancer origin, and thus normally silent, may become reactivated as a result of translocation bringing the corresponding region far from the nuclear lamina. The opposite situation, when a previously active gene is silenced due to the association of the corresponding locus to the lamina can also take place. In collaboration with B. van Steensel lab (NKI, Netherlands), we plan to investigate this question by studying the changes in epigenetic profiles and gene expression in genes located in the regions surrounding large structural variants in different cancer types.

To summarize, analysis of novel types of HTS data, application of HTS to new species, cancer, characterization of signal coming from the repetitive elements of the genome and analysis of single cell datasets, all this needs the development of novel specific data processing algorithms. The laboratory of Bioinformatics, Biostatistics, Epidemiology and computational Systems Biology of Cancer in Institut Curie I work in provides plenty of opportunities for collaboration with clinicians and biologists to study these questions. My ambition is to develop novel computational techniques for analysis of cancer HTS data and in parallel contribute to our understanding of cancer genetics and its connection with tumor aggressiveness and response to treatment. By using large scale datasets including clinical information as well as information obtained from HTS data (e.g. genetic, epigenetic, transcription profiles), I plan to develop new data analysis strategies for identification of key events disabling or amplifying signaling pathways in a given cancer. My work will contribute to our understanding of the mechanisms of regulation in normal and cancer cells. I hope to provide explanations why certain changes at the level of the nucleotide sequence result in epigenetic changes and changes in gene expression and as consequence stimulate tumor development or progression.